\documentstyle[eqsecnum,aps,epsf,
feynmf]{revtex}
\newcommand{\scs}{\scriptstyle}
\newcommand{\meas}{{\cal D}\bar{\psi}{\cal D}\psi{\cal D}A\,}
\newcommand{\qedaction}{{\cal A}_{\rm QED}[\bar{\psi},\psi,A]}

\newcommand{\threeint}{\int\!\!\int\!\!\int}
\newcommand{\twoint}{\int\!\!\int}
\newcommand{\freemean}[1]{\langle\,#1\,\rangle^{(0)}}
\newcommand{\mean}[1]{\langle\,#1\,\rangle}
\newcommand{\meanJ}[1]{\langle\,#1\,\rangle^J}
\newcommand{\trace}{{\rm Tr\,}}
\newcommand{\setval}{\fmfset{wiggly_len}{1.5mm}\fmfset{arrow_len}{1.5mm}
\fmfset{arrow_ang}{13}\fmfset{dash_len}{1.5mm}\fmfpen{0.125mm}
\fmfset{dot_size}{0.8thick}}
\newcommand{\dbose}[3]{\frac{\delta #1}{\delta
\parbox{10mm}{\centerline{
\begin{fmfgraph*}(5,3)
\setval
\fmfpen{0.125mm}
\fmfleft{v1}
\fmfright{v2}
\fmf{boson}{v2,v1}
\fmfv{decor.size=0,label={\footnotesize #2},l.dist=0.5mm}{v1}
\fmfv{decor.size=0,label={\footnotesize #3},l.dist=0.5mm}{v2}
\end{fmfgraph*}
}}}}
\newcommand{\dfermi}[3]{\frac{\delta #1}{\delta
\parbox{10mm}{\centerline{
\begin{fmfgraph*}(5,3)
\fmfset{arrow_len}{1.5mm}
\fmfpen{0.125mm}
\fmfleft{v1}
\fmfright{v2}
\fmf{fermion}{v2,v1}
\fmfv{decor.size=0,label={\footnotesize #2},l.dist=0.5mm}{v1}
\fmfv{decor.size=0,label={\footnotesize #3},l.dist=0.5mm}{v2}
\end{fmfgraph*}
}}}}
\newcommand{\ddfermi}[1]{\frac{\delta^2 #1}{\delta
\parbox{10mm}{\centerline{
\begin{fmfgraph*}(5,3)
\fmfset{arrow_len}{1.5mm}
\fmfpen{0.125mm}
\fmfleft{v1}
\fmfright{v2}
\fmf{fermion}{v2,v1}
\fmfv{decor.size=0,label={\footnotesize 1},l.dist=0.5mm}{v1}
\fmfv{decor.size=0,label={\footnotesize 2},l.dist=0.5mm}{v2}
\end{fmfgraph*}
}}\,\delta
\parbox{10mm}{\centerline{
\begin{fmfgraph*}(5,3)
\fmfset{arrow_len}{1.5mm}
\fmfpen{0.125mm}
\fmfleft{v1}
\fmfright{v2}
\fmf{fermion}{v2,v1}
\fmfv{decor.size=0,label={\footnotesize 3},l.dist=0.5mm}{v1}
\fmfv{decor.size=0,label={\footnotesize 4},l.dist=0.5mm}{v2}
\end{fmfgraph*}
}}
}}
\newcommand{\dvertex}[4]{\frac{\delta #1}{\rule[0pt]{0pt}{15pt}\delta
\parbox{10mm}{\centerline{
\begin{fmfgraph*}(4,3.464)
\fmfset{arrow_len}{1.5mm}
\fmfset{wiggly_len}{1.5mm}
\fmfset{dot_size}{0.7thick}
\fmfset{arrow_ang}{13}
\fmfpen{0.125mm}
\fmfforce{1w,0h}{v1}
\fmfforce{0w,0h}{v2}
\fmfforce{0.5w,1h}{v3}
\fmfforce{0.5w,0.2886h}{vm}
\fmf{fermion}{v1,vm,v2}
\fmf{boson}{v3,vm}
\fmfv{decor.size=0,label={\footnotesize #2},l.dist=0.5mm}{v1}
\fmfv{decor.size=0,label={\footnotesize #3},l.dist=0.5mm}{v2}
\fmfv{decor.size=0,label={\footnotesize #4},l.dist=0.5mm}{v3}
\fmfdot{vm}
\end{fmfgraph*}
}}\rule[0pt]{0pt}{15pt}}}
\newcommand{\dcurr}[2]{\frac{\delta #1}{\rule[0pt]{0pt}{5pt}\delta
\parbox{7mm}{\centerline{
\begin{fmfgraph*}(4,3.464)
\fmfset{arrow_len}{1.5mm}
\fmfset{wiggly_len}{1.5mm}
\fmfset{dot_size}{0.7thick}
\fmfset{arrow_ang}{13}
\fmfpen{0.125mm}
\fmfforce{0w,0h}{s1}
\fmfforce{0.5w,0.5h}{s2}
\fmfforce{0w,1h}{s3}
\fmfforce{1w,0.5h}{v}
\fmf{double,width=0.2mm}{s1,s2,s3}
\fmf{boson}{s2,v}
\fmfv{decor.size=0,label={\footnotesize #2},l.dist=0.5mm}{v}
\fmfdot{s2}
\end{fmfgraph*}
}}\rule[0pt]{0pt}{5pt}}}
\begin{document}
\setlength{\unitlength}{1mm}
\begin{fmffile}{diag1bkpB}
\title{
Recursive Graphical Construction of \\
Feynman Diagrams in Quantum Electrodynamics}
\author{Michael Bachmann, Hagen Kleinert, and Axel Pelster}
\address{Institut f\"ur Theoretische Physik, Freie Universit\"at Berlin, 
Arnimallee 14, 14195 Berlin, Germany}
\date{\today}
\maketitle
\begin{abstract}
We present a method for a recursive graphical construction of 
Feynman diagrams with their correct multiplicities
in quantum electrodynamics.
The method is first applied to find all diagrams contributing
to the vacuum energy 
from which all $n$-point functions are derived by functional
differentiation with respect to electron and photon 
propagators, and to the interaction.
Basis for our construction is a functional differential equation obeyed 
by the vacuum energy when considered as 
a functional of the free propagators and the interaction. 
Our method does not employ external sources in contrast to traditional
approaches.
\end{abstract}
\section{Introduction}
In quantum field theory, it is well known \cite{Streater,Schwinger}
that the complete knowledge of
all vacuum diagrams implies the knowledge of the entire theory
(``the vacuum is the world''). 
Indeed, it is possible to derive all correlation functions and scattering
amplitudes from the vacuum diagrams. This has been elaborated explicitly for
$\phi^4$ theory in the disordered phase in Refs.~\cite{phi4,Verena} and for the ordered 
phase in Ref.~\cite{Boris}, following a general theoretical 
framework laid out some time ago~\cite{Kleinert1,Kleinert2}. However,
this knowledge has
not yet been applied for constructing
an efficient algebraic method along these lines
for field theories of fundamental
particles.
The purpose of the present paper is to do this 
for quantum electrodynamics (QED). We show how to derive
systematically all Feynman diagrams of the theory together with their 
correct multiplicities in a two step process:
First we find the vacuum energy from a sum over all vacuum diagrams by a
recursive graphical procedure. This is developed by solving a functional
differential equation which involves functional
derivatives with respect to the free
electron and photon
propagators. In a second step, we find all correlation functions by a
diagrammatic application of  
functional derivatives upon the
vacuum energy. In contrast to conventional procedures
\cite{Drell,Amit,Zuber,Bellac,Zinn,Peskin},
no external currents coupled to single
fields are used, such that there is no need for 
Grassmann sources for the electron fields.
An additional advantage of our procedure is that the number 
of derivatives to be performed
for a certain correlation function is half as big as with external 
sources.

In Section~\ref{gentheo} we establish the
partition function of euclidean QED as a functional with respect to 
the inverse
electron and photon progators as well as a generalized interaction.
By setting up graphical representations for functional derivatives with
respect to these bilocal and trilocal functions, we show in Section~\ref{pert} 
that the partition function constitutes a generating functional
for all correlation functions. This forms the basis for a perturbative
expansion of the vacuum energy in terms of connected vacuum diagrams.
In Section~\ref{recrel} we then derive a recursion relation which allows
to graphically construct the connected vacuum diagrams order by order.
From these we obtain in Section~\ref{npoint} all diagrams for self interactions
and scattering processes 
by cutting electron as well as photon lines or by removing vertices.
Along similar lines we apply in Section~\ref{scatj} 
our method for scattering processes
in the presence of an external electromagnetic field.
\section{Generating Functional Without Particle Sources}
\label{gentheo}
We begin by setting up a generating functional for all Feynman diagrams
of quantum electrodynamics which 
does not employ external particle 
sources coupled linearly to the fields.
\subsection{Partition Function of QED}
Our notation for the action of QED in euclidean spacetime with
a gauge fixing of Feynman type is
\begin{equation}
  \label{gt01}
  {\cal A}_{\rm QED}[\bar{\psi},\psi,A]=\int d^4 x\,\left[\bar{\psi}_\alpha(i
\gamma^\mu_{\alpha\beta} 
\partial_\mu+m)\psi_\beta
+ \frac{1}{2} A_{\mu} ( - \partial^2 ) A^{\mu}  
- e \bar{\psi}_\alpha
\gamma^\mu_{\alpha\beta} A_\mu\psi_\beta\right]
\end{equation}
with Dirac spinor fields $\psi_\alpha,\bar{\psi}_\beta=
\psi_\alpha^\dagger\gamma^0_{\alpha\beta}$ ($\alpha,\beta=1,\ldots,4$) 
and Maxwell's vector field $A_\mu$ ($\mu=0,\ldots,3$).
The properties of
the vacuum are completely described by the partition function
\begin{equation}
  \label{gt02}
  Z_{\rm QED}=\oint\meas {\rm e}^{-\qedaction},
\end{equation}
where the electron fields $\bar{\psi}$ and $\psi$ are Grassmannian.
Let us split the action into the three terms
\begin{equation}
  \label{gt03}
  \qedaction={\cal A}_{\psi}[\bar{\psi},\psi]+{\cal A}_{A}[A]+
{\cal A}_{\rm int}[\bar{\psi},\psi,A],
\end{equation}
corresponding to the Dirac, Maxwell,
and interaction terms in (\ref{gt01}).
For the upcoming development it will be useful to consider 
the free parts of the action 
as bilocal functionals. The free action of the electrons
is 
\begin{equation}
  \label{gt04}
  {\cal A}_{\psi}[\bar{\psi},\psi]=\twoint d^4x d^4x'
\,\bar{\psi}_\alpha(x)\,
S^{-1}_{{\rm F}\alpha\beta}(x,x')\,\psi_\beta(x'),
\end{equation}
with a kernel
\begin{equation}
  \label{gt05}
  S^{-1}_{{\rm F}\alpha\beta}(x,x')=(i\gamma_{\alpha\beta}^\mu
\partial_\mu+m\delta_{\alpha\beta})\delta(x-x'),
\end{equation}
while the free action for the Maxwell field reads
\begin{equation}
  \label{gt06}
  {\cal A}_{A}[A]=\frac{1}{2}\twoint d^4x d^4x'\,A^\mu(x)\,
D^{-1}_{{\rm F}\mu\nu}(x,x')\,A^\nu(x')
\end{equation}
with a kernel
\begin{equation}
  \label{gt07}
  D^{-1}_{{\rm F}\mu\nu}(x,x')=-\partial^2 \delta(x-x')\delta_{\mu\nu}.
\end{equation}
In the following, we shall omit all 
vector and spinor indices, for brevity.
\subsection{Generalized Action}
Our generating functional will arise from a generalization  
of the free action
\begin{equation}
\label{gt11}
{\cal A}^{(0)}[\bar{\psi},\psi,A]={\cal A}_{\psi}[\bar{\psi},\psi]+
{\cal A}_{A}[A]
\end{equation}
to bilocal functionals of arbitrary kernels $S^{-1} ( x_1 , x_2 )$
and $D^{-1} ( x_1 , x_2 )=D^{-1} ( x_2 , x_1 )$ according to
\begin{eqnarray}
  \label{gt16}
  {\cal A}_{\psi}[\bar{\psi},\psi]\;\to\;
{\cal A}_{\psi}[\bar{\psi},\psi;S^{-1}]&=&\twoint d^4x_1d^4x_2\,
\bar{\psi}(x_1)\,S^{-1}(x_1,x_2)\,\psi(x_2),\\
  \label{gt17}
  {\cal A}_A[A]\;\to\;{\cal A}_{A}[A;D^{-1}]&=&
\frac{1}{2}\twoint d^4x_1d^4x_2\,A(x_1)\,
D^{-1}(x_1,x_2)\,A(x_2).
\end{eqnarray}
The kernels $S^{-1} ( x_1 , x_2 )$
and $D^{-1} ( x_1 , x_2 )$
are only required to 
possess 
a functional
inverse $S(x_1,x_2)$ and $D(x_1,x_2)$.
Similarly, we shall generalize the interaction to
\begin{equation}
  \label{gt09}
{\cal A}_{\rm int}[\bar{\psi},\psi,A]\;\to\;  
{\cal A}_{\rm int}[\bar{\psi},\psi,A;V]=- e \threeint d^4x_1d^4x_2d^4x_3\,
V(x_1,x_2;x_3)\,\bar{\psi}(x_1)\psi(x_2) A(x_3),
\end{equation}
where $V(x_1,x_2;x_3)$ is an arbitrary trilocal function.
At the end we shall return to QED
by substituting $S\to S_{\rm F}, D\to D_{\rm F}$ and 
$e V(x_1,x_2;x_3)\rightarrow 
e \gamma_\mu \delta(x_1-x_2) \delta(x_1-x_3)$.

The generalized partition function
\begin{equation}
  \label{gt18}
  Z=\oint\meas\,{\rm e}^{-{\cal A}[\bar{\psi},\psi,A;S^{-1},D^{-1},V]}
\end{equation}
with the action
\begin{equation}
  \label{gt18b}
  {\cal A}[\bar{\psi},\psi,A;S^{-1},D^{-1},V]=
{\cal A}_{\psi}[\bar{\psi},\psi;S^{-1}]+
{\cal A}_{A}[\bar{\psi},\psi;D^{-1}]+
{\cal A}_{\rm int}[\bar{\psi},\psi,A;V]
\end{equation}
then represents
a functional of the bilocal quantities $S^{-1}(x_1,x_2),
D^{-1}(x_1,x_2)$, and of 
the trilocal function $V(x_1,x_2;x_3)$.
All $n$-point correlation functions of the theory are obtained from
expectation values defined by
\begin{equation}
  \label{gt18c}
  \mean{\hat{O}_1(x_1)\,\hat{O}_2(x_2)\,\cdots}=Z^{-1}
\oint\meas\,O_1(x_1)O_2(x_2)\cdots {\rm e}^{-{\cal A}[\bar{\psi},\psi,A;S^{-1},
D^{-1},V]} ,
\end{equation}
where
the local operators $\hat{O}_i(x)$ are products of field operators 
$\hat{\psi}(x)$, $\hat{\bar{\psi}}(x)$, and $\hat{A}(x)$ at the same 
spacetime point.
Important examples for expectation values of this kind are the 
photon and the electron propagators of the interacting theory
\begin{eqnarray}
\label{gt18d}
{}^\gamma G^{2}(x_1,x_2)&\equiv&\mean{\hat{A}(x_1)\,\hat{A}(x_2)}
=Z^{-1}
\oint\meas\,A(x_1)A(x_2)\,{\rm e}^{-{\cal A}[\bar{\psi},\psi,A;S^{-1},
D^{-1},V]},\\
\label{gt18e}
{}^e G^{2}(x_1,x_2)&\equiv&\mean{\hat{\psi}(x_1)\,\hat{\bar{\psi}}(x_2)}
=Z^{-1}
\oint\meas\,\psi(x_1)\bar{\psi}(x_2)\,{\rm e}^{-{\cal A}[\bar{\psi},
\psi,A;S^{-1},
D^{-1},V]}.
\end{eqnarray}
For a perturbative calculation of the partition function $Z$ we define
the free vacuum functional
\begin{equation}
  \label{gt10}
  Z^{(0)}\equiv\oint\meas\,{\rm e}^{-{\cal A}^{(0)}[\bar{\psi},\psi,A;
S^{-1},D^{-1}]},
\end{equation}
whose action is quadratic in the fields. The path integral is Gaussian 
and yields
\begin{equation}
  \label{gt19}
  Z^{(0)}=\exp\left[\trace {\rm ln}\,S^{-1}\right]\,\exp\left[-\frac{1}{2}
\trace {\rm ln}\,D^{-1} \right].
\end{equation}
The free correlation functions
of arbitrary local electron and photon operators 
$\hat{O}(x)$ are defined by the free part of the expectation
values (\ref{gt18c})
\begin{equation}
  \label{gt12}
  \freemean{\hat{O}_1(x_1)\,\hat{O}_2(x_2)\,\cdots}=[Z^{(0)}]^{-1}
\oint\meas\,O_1(x_1)O_2(x_2)\cdots {\rm e}^{-{\cal A}^{(0)}[\bar{\psi},\psi,A;
S^{-1},D^{-1}]},
\end{equation}
and the free-field propagators are the expectation values
\begin{eqnarray}
  \label{gt15}
{}^\gamma G^{2^{(0)}}(x_1,x_2) =
D(x_1,x_2)&\equiv&\freemean{\hat{A}(x_1)\,\hat{A}(x_2)}
\equiv D(x_2,x_1),\\
\label{gt14}
 {}^e G^{2^{(0)}}(x_1,x_2) = S(x_1,x_2)&\equiv&\freemean{\hat{\psi}(x_1)\,
\hat{\bar{\psi}}(x_2)}.
\end{eqnarray}
To avoid a pile up of infinite volume factors in a perturbation
expansion, 
it is favourable to 
go over from $Z^{(0)}$ to the negative vacuum energy $W^{(0)}$ defined by
\begin{equation}
  \label{gt19b}
  W^{(0)}\equiv{\rm ln}\,Z^{(0)}=W_{\psi}^{(0)}+W_{A}^{(0)},
\end{equation}
where the free electron and photon parts are 
\begin{equation}
  \label{gt19d}
  W_{\psi}^{(0)}=\trace {\rm ln}\,S^{-1}
\end{equation}
and
\begin{equation}
  \label{gt19c}
  W_{A}^{(0)}=-\frac{1}{2}\trace {\rm ln}\,D^{-1}.
\end{equation}
The negative total vacuum energy
\begin{equation}
  \label{gt19e}
  W={\rm ln}\,Z
\end{equation}
is obtained perturbatively by expanding the functional integral 
(\ref{gt18}) in powers of the coupling constant $e$: 
\begin{eqnarray}
\label{gt19f}
W = \sum_{p=0}^{\infty} \, e^{2p} \, W^{(p)},
\end{eqnarray}
where the quantities $W^{(p)}$ with $p\ge 1$ are free-field 
expectation values 
of the type (\ref{gt12}):
\begin{equation}
  \label{gt19g}
  W^{(p)}=\int_{1\cdots 6p}V_{123}\cdots V_{6p-2\,6p-1\,6p}
\freemean{\hat{\psi}_{6p-1}\cdots\hat{\psi}_{5}\,\hat{\psi}_{2}\,
\hat{\bar{\psi}}_1 \,\hat{\bar{\psi}}_4
\cdots\hat{\bar{\psi}}_{6p-2}\,\hat{A}_{3}\,\hat{A}_{6}
\cdots\hat{A}_{6p}},\quad p\ge 1.
\end{equation}
Throughout this paper we shall use from now on the short-hand notation 
$1=x_1,2=x_2,\ldots$ and 
$\int_{1\cdots}=\int d^4x_1\cdots$ .
The expectation values in (\ref{gt19g}) are evaluated with the help of 
Wick's rule as a
sum of Feynman integrals, which are pictured
as connected vacuum diagrams 
constructed from lines and vertices. A straight line with an arrow 
represents
an electron propagator
\begin{eqnarray}
\label{vac07}
\parbox{20mm}{\centerline{
\begin{fmfgraph*}(7,3)
\setval
\fmfleft{v1}
\fmfright{v2}
\fmf{fermion}{v2,v1}
\fmflabel{${\scs 1}$}{v1}
\fmflabel{${\scs 2}$}{v2}
\end{fmfgraph*}}}  
\equiv \quad S_{12},
\end{eqnarray}
whereas a wiggly line stands for a photon propagator
\begin{eqnarray}
\label{vac08}
\parbox{20mm}{\centerline{
\begin{fmfgraph*}(7,3)
\setval
\fmfleft{v1}
\fmfright{v2}
\fmf{boson}{v1,v2}
\fmflabel{${\scs 1}$}{v1}
\fmflabel{${\scs 2}$}{v2}
\end{fmfgraph*}
}}\equiv \quad D_{12}.
\end{eqnarray}
The vertex represents an integral over the interaction
potential:
\begin{eqnarray}
\label{vac09}
\parbox{15mm}{\centerline{
\begin{fmfgraph}(5,4.33)
\setval
\fmfforce{1w,0h}{v1}
\fmfforce{0w,0h}{v2}
\fmfforce{0.5w,1h}{v3}
\fmfforce{0.5w,0.2886h}{vm}
\fmf{fermion}{v1,vm,v2}
\fmf{boson}{v3,vm}
\fmfdot{vm}
\end{fmfgraph}
}}\equiv \quad e \int_{123} V_{123}.
\end{eqnarray}
The vacuum energies (\ref{gt19d}) and (\ref{gt19c}) will be represented by 
single-loop diagrams 
\begin{equation}
  \label{vac09b}
  W_{\psi}^{(0)}= -\;
\parbox{8mm}{\centerline{
\begin{fmfgraph}(5,5)
\setval
\fmfi{fermion}{reverse fullcircle scaled 1w shifted (0.5w,0.5h)}
\end{fmfgraph}
}}
\end{equation}
and
\begin{equation}
  \label{vac09c}
  W_{A}^{(0)}=\frac{1}{2}\;
\parbox{8mm}{\centerline{
\begin{fmfgraph}(5,5)
\setval
\fmfi{boson}{reverse fullcircle scaled 1w shifted (0.5w,0.5h)}
\end{fmfgraph}
}}.
\end{equation}
This leaves us with the important problem of finding
all connected vacuum diagrams. For this we shall exploit that 
the partition function 
(\ref{gt18}) is a functional of the bilocal 
functions $S^{-1}(x_1,x_2)$, 
$D^{-1}(x_1,x_2)$, and of the trilocal function $V (x_1,x_2;x_3)$.
\section{Perturbation Theory}
\label{pert}
As a preparation for our generation procedure for vacuum diagrams,
we set up a graphical representation of
functional
derivatives with respect to the kernels $S^{-1}$, $D^{-1}$, the propagators
$S$, $D$, and the interaction function $V$. 
After this we express the vacuum functional $W$ in terms of
a series of functional derivatives of the free partition function
$Z^{(0)}$ with respect to the kernels. 
\subsection{Functional Derivatives With Respect to $S^{-1}(x_1,x_2)$,
$D^{-1}(x_1,x_2)$, and $V(x_1,x_2;x_3)$}
\label{funcderiv}
Each Feynman diagram is composed of integrals over products of the propagators
$S$, $D$ and may thus be considered as a functional of the kernels $S^{-1}$,
$D^{-1}$. In the following we set up the graphical rules for performing
functional derivatives with respect
to these functional matrices. With these rules we can
generate all $2n$-point correlation functions with $n=1,2,\ldots$
from vacuum diagrams.
To produce also ($2n+1$)-point correlation functions  with $n=1,2,\ldots$ 
such as the fundamental three-point vertex function from vacuum diagrams, 
it is useful to introduce additionally
a functional derivative with respect to the interaction function $V_{123}$.
\subsubsection{Functional Derivative with Respect to the Photon Kernel}
The kernel $D^{-1}_{12}$ of the photon is symmetric $D^{-1}_{12}=D^{-1}_{21}$,
so that the basic functional derivatives are also symmetric \cite{phi4,Verena}
\begin{eqnarray}
\label{rr07}
\frac{\delta D^{-1}_{12}}{\delta D^{-1}_{34}}\equiv \frac{1}{2}
\left\{\delta_{13}\delta_{42}+\delta_{14}\delta_{32}
\right\},
\end{eqnarray}
as is discussed in detail in Ref.~\cite{Boris}.
By the chain rule of differentiation, this defines the functional derivative
with respect to $D^{-1}$ for all functionals of $D^{-1}$.
As an example, we calculate 
the free photon propagator (\ref{gt15}) by 
applying the operator $\delta/\delta D^{-1}_{12}$ to Eq. (\ref{gt10}).
Taking into account Eq.~(\ref{gt19b}) and Eq.~(\ref{rr07}), we find
\begin{eqnarray}
\label{rr09}
D_{12}=-2\,\frac{\delta W_{A}^{(0)}}{\delta D^{-1}_{12}}.
\end{eqnarray}
Inserting the explicit form (\ref{gt19c}), we obtain 
\begin{equation}
  \label{rr11}
  D_{12}=\frac{\delta}{\delta D^{-1}_{12}}\trace\,{\rm ln}\,D^{-1},
\end{equation}
With the notation (\ref{vac08}) and (\ref{vac09c}), we can write relation 
(\ref{rr09}) graphically as
\begin{equation}
\label{rr12b}
\parbox{20mm}{\centerline{
\begin{fmfgraph*}(7,3)
\setval
\fmfleft{v1}
\fmfright{v2}
\fmf{boson}{v1,v2}
\fmflabel{$\scs{1}$}{v1}
\fmflabel{$\scs{2}$}{v2}
\end{fmfgraph*}
}}
=-\frac{\delta}{\delta D^{-1}_{12}}\;
\parbox{8mm}{\centerline{
\begin{fmfgraph}(5,5)
\setval
\fmfi{boson}{reverse fullcircle scaled 1w shifted (0.5w,0.5h)}
\end{fmfgraph}
}}.
\end{equation}
This diagrammatic equation may be viewed as a special case of a general 
graphical rule derived as follows:
Let us apply the functional 
derivative (\ref{rr07}) to a photon propagator $D_{12}$. Because of
the identity
\begin{equation}
  \label{rr14}
  \int_{\bar{1}}\,D_{1\bar{1}} D^{-1}_{\bar{1}2}=\delta_{12}
\end{equation}
we find
\begin{equation}
  \label{rr15}
  -\frac{\delta D_{12}}{\delta D^{-1}_{34}}=\frac{1}{2}
\left\{D_{13}D_{42}+D_{14}D_{32}\right\}.
\end{equation}
Diagrammatically, this equation implies that
the operation $-\delta/\delta D^{-1}_{34}$ applied to a photon line 
(\ref{vac08}) amounts to cutting the line:
\begin{equation}
\label{rr13}
-\,\frac{\delta}{\delta D^{-1}_{34}}
\parbox{20mm}{\centerline{
\begin{fmfgraph*}(7,3)
\setval
\fmfleft{v1}
\fmfright{v2}
\fmf{boson}{v1,v2}
\fmflabel{$\scs{1}$}{v1}
\fmflabel{$\scs{2}$}{v2}
\end{fmfgraph*}
}} =\frac{1}{2}\,\Bigg\{
\parbox{20mm}{\centerline{
\begin{fmfgraph*}(7,3)
\setval
\fmfleft{v1}
\fmfright{v2}
\fmf{boson}{v1,v2}
\fmflabel{$\scs{1}$}{v1}
\fmflabel{$\scs{3}$}{v2}
\end{fmfgraph*}
}}
\parbox{20mm}{\centerline{
\begin{fmfgraph*}(7,3)
\setval
\fmfleft{v1}
\fmfright{v2}
\fmf{boson}{v1,v2}
\fmflabel{$\scs{4}$}{v1}
\fmflabel{$\scs{2}$}{v2}
\end{fmfgraph*}
}} +
\parbox{20mm}{\centerline{
\begin{fmfgraph*}(7,3)
\setval
\fmfleft{v1}
\fmfright{v2}
\fmf{boson}{v1,v2}
\fmflabel{$\scs{1}$}{v1}
\fmflabel{$\scs{4}$}{v2}
\end{fmfgraph*}
}}
\parbox{20mm}{\centerline{
\begin{fmfgraph*}(7,3)
\setval
\fmfleft{v1}
\fmfright{v2}
\fmf{boson}{v1,v2}
\fmflabel{$\scs{3}$}{v1}
\fmflabel{$\scs{2}$}{v2}
\end{fmfgraph*}
}}\Bigg\}.
\end{equation}
Note that the indices of the kernel $D_{34}^{-1}$ are symmetrically 
attached to the newly created line ends in the two possible ways due
to the differentiation rule (\ref{rr07}). 
This rule implies directly the diagrammatic equation (\ref{rr12b}).

Consider now higher-order correlation functions which follow from higher
functional derivatives of $W^{(0)}_A$. From the definition 
(\ref{gt12}) and Eq.~(\ref{gt17}), we 
obtain the free four-point function as the second functional derivative
\begin{equation}
  \label{rr16}
  {}^\gamma G^{{4}^{(0)}}_{1234}\equiv\freemean{\hat{A}_1\,
\hat{A}_2\,\hat{A}_3\,\hat{A}_4}=4\,{\rm e}^{-W_{A}^{(0)}}\,
\frac{\delta^2}{\delta D^{-1}_{12} \delta D^{-1}_{34}}{\rm e}^{W_{A}^{(0)}}.
\end{equation}
Because of the symmetry of $D_{12}$,
the order in which the spacetime arguments appear in the inverse 
propagators is of no importance. 
Inserting for $W^{(0)}_A$ the explicit form (\ref{gt19c}),
the first derivative yields via Eq.~(\ref{rr11}) 
just $-D_{34}\exp\{W_{A}^{(0)}\}$, the second derivative 
applied to this gives with the rule (\ref{rr15}) and, once more (\ref{rr11}), 
\begin{equation}
  \label{rr17}
  {}^\gamma G^{{4}^{(0)}}_{1234}=D_{13}D_{24}+D_{32}D_{14}+
D_{12}D_{34}.
\end{equation}
The right-hand side has the graphical representation
\begin{equation}
  \label{rr18}
 {}^\gamma G^{{4}^{(0)}}_{1234} =
\parbox{22mm}{\begin{center}
\begin{fmfgraph*}(8,5)
\setval
\fmfleft{v3,v1}
\fmfright{v4,v2}
\fmf{boson}{v1,v4}
\fmf{boson}{v3,v2}
\fmflabel{$\scs{2}$}{v1}
\fmflabel{$\scs{3}$}{v2}
\fmflabel{$\scs{1}$}{v3}
\fmflabel{$\scs{4}$}{v4}
\end{fmfgraph*}
\end{center}}+
\parbox{22mm}{\begin{center}
\begin{fmfgraph*}(8,5)
\setval
\fmfleft{v3,v1}
\fmfright{v4,v2}
\fmf{boson}{v1,v2}
\fmf{boson}{v3,v4}
\fmflabel{$\scs{2}$}{v1}
\fmflabel{$\scs{3}$}{v2}
\fmflabel{$\scs{1}$}{v3}
\fmflabel{$\scs{4}$}{v4}
\end{fmfgraph*}
\end{center}}+
\parbox{22mm}{\begin{center}
\begin{fmfgraph*}(8,5)
\setval
\fmfleft{v3,v1}
\fmfright{v4,v2}
\fmf{boson}{v1,v3}
\fmf{boson}{v2,v4}
\fmflabel{$\scs{2}$}{v1}
\fmflabel{$\scs{3}$}{v2}
\fmflabel{$\scs{1}$}{v3}
\fmflabel{$\scs{4}$}{v4}
\end{fmfgraph*}
\end{center}}.
\end{equation}
The same diagrams are obtained by applying the cutting rule (\ref{rr13})
twice to the single-loop diagram (\ref{vac09c}).

While derivatives with respect to the kernel $D^{-1}$ amount  to cutting
photon lines, we show now that
derivatives with respect to the photon propagator $D$ 
amount to line amputations.
The transformation rule between the two operations
follows from relation (\ref{rr15}):
\begin{equation}
  \label{rr19}
  \frac{\delta}{\delta D^{-1}_{12}}=-\int_{34}\,D_{13} 
D_{24} \frac{\delta}{\delta D_{34}},
\end{equation}
which is equivalent to
\begin{equation}
  \label{rr19b}
  \frac{\delta}{\delta D_{12}}=-\int_{34}\,D^{-1}_{13} 
D^{-1}_{24} \frac{\delta}{\delta D^{-1}_{34}}.
\end{equation}
The functional derivative with respect to $D_{12}$ satisfies of course
the fundamental relation (\ref{rr07}):
\begin{equation}
  \label{rr19c}
  \frac{\delta D_{12}}{\delta D_{34}}=\frac{1}{2}\left\{
\delta_{13}\delta_{42}+\delta_{14}\delta_{32} \right\}.
\end{equation}
We shall represent the right-hand side graphically by
extending the Feynman diagrams by the symbol:
\begin{eqnarray}
\label{rr19d}
{\scs 1}
\parbox{6mm}{\centerline{
\begin{fmfgraph}(4,3)
\setval
\fmfforce{0w,0.5h}{i1}
\fmfforce{1w,0.5h}{o1}
\fmfforce{0.5w,0.5h}{v1}
\fmf{boson}{i1,v1}
\fmf{boson}{v1,o1}
\fmfv{decor.size=0}{i1}
\fmfv{decor.size=0}{o1}
\fmfv{decor.shape=circle,decor.filled=empty,decor.size=0.6mm}{v1}
\end{fmfgraph}}} {\scs 2}
\quad &\equiv& \quad \delta_{12}.
\end{eqnarray}
If we write the functional derivative with respect to the propagator 
$D_{12}$ graphically as
\begin{eqnarray}
\label{rr19e}
\frac{\delta}{\delta D_{12}} \quad &\equiv&\quad\dbose{}{1}{2},
\end{eqnarray}
we may express Eq.~(\ref{rr19c}) as
\begin{eqnarray}
\dbose{}{1}{2} 
\parbox{20mm}{\centerline{
\begin{fmfgraph*}(7,3)
\setval
\fmfleft{v1}
\fmfright{v2}
\fmf{boson}{v1,v2}
\fmflabel{${\scs 3}$}{v1}
\fmflabel{${\scs 4}$}{v2}
\end{fmfgraph*}
}} = \frac{1}{2} \,\Bigg\{
\quad{\scs 1}
\parbox{6mm}{\centerline{
\begin{fmfgraph}(4,3)
\setval
\fmfforce{0w,0.5h}{i1}
\fmfforce{1w,0.5h}{o1}
\fmfforce{0.5w,0.5h}{v1}
\fmf{boson}{i1,v1}
\fmf{boson}{v1,o1}
\fmfv{decor.size=0}{i1}
\fmfv{decor.size=0}{o1}
\fmfv{decor.shape=circle,decor.filled=empty,decor.size=0.6mm}{v1}
\end{fmfgraph}}} {\scs 3} \quad
{\scs 4}
\parbox{6mm}{\centerline{
\begin{fmfgraph}(4,3)
\setval
\fmfforce{0w,0.5h}{i1}
\fmfforce{1w,0.5h}{o1}
\fmfforce{0.5w,0.5h}{v1}
\fmf{boson}{i1,v1}
\fmf{boson}{v1,o1}
\fmfv{decor.size=0}{i1}
\fmfv{decor.size=0}{o1}
\fmfv{decor.shape=circle,decor.filled=empty,decor.size=0.6mm}{v1}
\end{fmfgraph}}} {\scs 2} \quad + \quad
{\scs 1}
\parbox{6mm}{\centerline{
\begin{fmfgraph}(4,3)
\setval
\fmfforce{0w,0.5h}{i1}
\fmfforce{1w,0.5h}{o1}
\fmfforce{0.5w,0.5h}{v1}
\fmf{boson}{i1,v1}
\fmf{boson}{v1,o1}
\fmfv{decor.size=0}{i1}
\fmfv{decor.size=0}{o1}
\fmfv{decor.shape=circle,decor.filled=empty,decor.size=0.6mm}{v1}
\end{fmfgraph}}} {\scs 4} \quad
{\scs 3}
\parbox{6mm}{\centerline{
\begin{fmfgraph}(4,3)
\setval
\fmfforce{0w,0.5h}{i1}
\fmfforce{1w,0.5h}{o1}
\fmfforce{0.5w,0.5h}{v1}
\fmf{boson}{i1,v1}
\fmf{boson}{v1,o1}
\fmfv{decor.size=0}{i1}
\fmfv{decor.size=0}{o1}
\fmfv{decor.shape=circle,decor.filled=empty,decor.size=0.6mm}{v1}
\end{fmfgraph}}} {\scs 2}\quad
\Bigg\}.
\end{eqnarray}
Thus, 
differentiating a photon line with respect to the corresponding propagator
amputates this line, leaving only the symmetrized indices at the end points. 
\subsubsection{Functional Derivative with Respect to the Electron Kernel}
Setting up graphical representations for functional derivatives for 
electrons is different from that in the 
photon case since the kernel $S^{-1}$ is no longer symmetric. The functional
derivative is therefore the usual one
\begin{equation}
  \label{rr20}
  \frac{\delta S^{-1}_{12}}{\delta S^{-1}_{34}}=\delta_{13}
\delta_{42},
\end{equation}
from which all others are derived via the chain rule of differentiation. 
The free electron propagator $S_{12}$ is found in analogy to (\ref{rr09}) by 
differentiating the free electron vacuum functional (\ref{gt19d}) with 
respect to the 
inverse electron propagator $S^{-1}$:  
\begin{equation}
  \label{rr21}
  S_{12}=\frac{\delta W_{\psi}^{(0)}}{\delta S^{-1}_{21}}.
\end{equation}
This implies
\begin{equation}
  \label{rr23}
  S_{12}=\frac{\delta}{\delta S^{-1}_{21}}\, \trace\,{\rm ln}\,S^{-1},
\end{equation}
which follows also from Eq.~(\ref{rr20}) and the chain rule of 
differentiation.
The graphical interpretation of the functional derivative 
$\delta/\delta S^{-1}_{21}$ is quite analogous to the photon case. 
In analogy to Eq.~(\ref{rr12b}), we write expression (\ref{rr23}) 
diagrammatically as
\begin{equation}
\label{rr24b}
\parbox{20mm}{\centerline{
\begin{fmfgraph*}(7,3)
\setval
\fmfleft{v1}
\fmfright{v2}
\fmf{fermion}{v2,v1}
\fmflabel{$\scs{1}$}{v1}
\fmflabel{$\scs{2}$}{v2}
\end{fmfgraph*}}}
=-\frac{\delta}{\delta S^{-1}_{21}}\;
\parbox{8mm}{\centerline{
\begin{fmfgraph}(5,5)
\setval
\fmfi{fermion}{reverse fullcircle scaled 1w shifted (0.5w,0.5h)}
\end{fmfgraph}
}}.
\end{equation}
This, in turn, can be understood as being a consequence of the general
cutting rule for electron lines:
\begin{equation}
\label{rr25}
\frac{\delta}{\delta S^{-1}_{43}}
\parbox{20mm}{\centerline{
\begin{fmfgraph*}(7,3)
\setval
\fmfleft{v1}
\fmfright{v2}
\fmf{fermion}{v2,v1}
\fmflabel{$\scs{1}$}{v1}
\fmflabel{$\scs{2}$}{v2}
\end{fmfgraph*}
}} =-
\parbox{22mm}{\begin{center}
\begin{fmfgraph*}(8,5)
\setval
\fmfleft{v1,v3}
\fmfright{v2,v4}
\fmfforce{0.5w,0.5h}{v5}
\fmf{plain}{v4,v5}
\fmf{plain}{v2,v5}
\fmf{fermion}{v5,v1}
\fmf{fermion}{v5,v3}
\fmflabel{$\scs{1}$}{v3}
\fmflabel{$\scs{3}$}{v1}
\fmflabel{$\scs{4}$}{v2}
\fmflabel{$\scs{2}$}{v4}
\end{fmfgraph*}
\end{center}},
\end{equation}
which graphically expresses the derivative relation
\begin{equation}
  \label{rr26}
  \frac{\delta S_{12}}{\delta S^{-1}_{43}}=-S_{14}S_{32}.
\end{equation}
The free electron 4-point function is obtained from two 
functional derivatives according to
\begin{equation}
  \label{rr27}
 {}^eG^{{4}^{(0)}}_{1234}\equiv\freemean{\hat{\psi}_1\,
\hat{\psi}_2\,\hat{\bar{\psi}}_3\,\hat{\bar{\psi}}_4}=
{\rm e}^{-W_{\psi}^{(0)}}\frac{\delta^2}{\delta S^{-1}_{32} \delta S^{-1}_{41}}
\,{\rm e}^{W_{\psi}^{(0)}}.
\end{equation}
Here, the electron fields must be properly rearranged to 
$\hat{\psi}_2\hat{\bar{\psi}}_3\hat{\psi}_1
\hat{\bar{\psi}}_4$ for applying the functional derivatives 
with respect to $S^{-1}$. Using Eqs.~(\ref{rr21}) and (\ref{rr26}) 
we obtain from Eq.~(\ref{rr27})
\begin{equation}
  \label{rr28}
  {}^eG^{{4}^{(0)}}_{1234}=S_{23}S_{14}-S_{24}S_{13},
\end{equation}
or graphically
\begin{equation}
  \label{rr29}
  {}^eG^{{4}^{(0)}}_{1234}=
  \parbox{22mm}{\begin{center}
\begin{fmfgraph*}(8,5)
\setval
\fmfleft{v1,v3}
\fmfright{v2,v4}
\fmf{fermion}{v2,v1}
\fmf{fermion}{v4,v3}
\fmflabel{$\scs{2}$}{v3}
\fmflabel{$\scs{1}$}{v1}
\fmflabel{$\scs{4}$}{v2}
\fmflabel{$\scs{3}$}{v4}
\end{fmfgraph*}
\end{center}}-
\parbox{22mm}{\begin{center}
\begin{fmfgraph*}(8,5)
\setval
\fmfleft{v1,v3}
\fmfright{v2,v4}
\fmfforce{0.5w,0.5h}{v5}
\fmf{plain}{v4,v5}
\fmf{plain}{v2,v5}
\fmf{fermion}{v5,v1}
\fmf{fermion}{v5,v3}
\fmflabel{$\scs{2}$}{v3}
\fmflabel{$\scs{1}$}{v1}
\fmflabel{$\scs{4}$}{v2}
\fmflabel{$\scs{3}$}{v4}
\end{fmfgraph*}
\end{center}}.
\end{equation}
Derivatives with respect to the propagators $S$ satisfy the relation
\begin{equation}
  \label{rr30}
  \frac{\delta S_{12}}{\delta S_{34}}=\delta_{13}\delta_{42},
\end{equation}
which in analogy to (\ref{rr19c}) is represented graphically as an 
amputation of an electron line
\begin{eqnarray}
\dfermi{}{3}{4} 
\parbox{20mm}{\centerline{
\begin{fmfgraph*}(7,3)
\setval
\fmfleft{v1}
\fmfright{v2}
\fmf{fermion}{v2,v1}
\fmflabel{${\scs 1}$}{v1}
\fmflabel{${\scs 2}$}{v2}
\end{fmfgraph*}
}} = 
{\scs 1}
\parbox{7mm}{\centerline{
\begin{fmfgraph}(5,3)
\setval
\fmfforce{0w,0.5h}{i1}
\fmfforce{1w,0.5h}{o1}
\fmfforce{0.5w,0.5h}{v1}
\fmf{fermion}{v1,i1}
\fmf{fermion}{o1,v1}
\fmfv{decor.size=0}{i1}
\fmfv{decor.size=0}{o1}
\fmfv{decor.shape=circle,decor.filled=empty,decor.size=0.6mm}{v1}
\end{fmfgraph}}} {\scs 3} \quad
{\scs 4}
\parbox{7mm}{\centerline{
\begin{fmfgraph}(5,3)
\setval
\fmfforce{0w,0.5h}{i1}
\fmfforce{1w,0.5h}{o1}
\fmfforce{0.5w,0.5h}{v1}
\fmf{fermion}{v1,i1}
\fmf{fermion}{o1,v1}
\fmfv{decor.size=0}{i1}
\fmfv{decor.size=0}{o1}
\fmfv{decor.shape=circle,decor.filled=empty,decor.size=0.6mm}{v1}
\end{fmfgraph}}} {\scs 2}.
\end{eqnarray}
Here we have introduced the additional diagrammatic symbols
\begin{eqnarray}
\label{rr30a}
{\scs 1}
\parbox{7mm}{\centerline{
\begin{fmfgraph}(5,3)
\setval
\fmfforce{0w,0.5h}{i1}
\fmfforce{1w,0.5h}{o1}
\fmfforce{0.5w,0.5h}{v1}
\fmf{fermion}{v1,i1}
\fmf{fermion}{o1,v1}
\fmfv{decor.size=0}{i1}
\fmfv{decor.size=0}{o1}
\fmfv{decor.shape=circle,decor.filled=empty,decor.size=0.6mm}{v1}
\end{fmfgraph}}} {\scs 2}
\quad &\equiv& \quad \delta_{12},\\
\label{rr30b}
\frac{\delta}{\delta S_{12}}\quad &\equiv&\quad\dfermi{}{1}{2}.
\end{eqnarray}
Differentiating an electron line with respect to the corresponding propagator
removes this line, leaving only the indices at the end points of the
remaining lines.

The analytic
relations between cutting and amputating lines are now, just as in 
Eqs.~(\ref{rr19}) and (\ref{rr19b}):
\begin{eqnarray}
  \label{rr31}
  \frac{\delta}{\delta S^{-1}_{12}}&=&-\int_{34} \,S_{31}
S_{24}\frac{\delta}{\delta S_{34}},\\
  \label{rr32}
  \frac{\delta}{\delta S_{12}}&=&-\int_{34} \,S^{-1}_{31}
S^{-1}_{24}\frac{\delta}{\delta S^{-1}_{34}}.
\end{eqnarray}
With the above graphical representations of the functional derivatives, it
will be possible to derive systematically all vacuum diagrams of the 
interacting theory order by 
order in the coupling strength $e$, and from these all diagrams
with an even number of legs.
\subsubsection{Functional Derivative with Respect to the Interaction}
If we want to find amplitudes involving an odd number of photons such as
the three-point function from vacuum diagrams, the
derivatives with respect to the kernels $S^{-1}$, $D^{-1}$ are not enough.
Here the general trilocal interaction function $V_{123}$ of
Eq.~(\ref{gt09}) is needed.
Thus,
we define an associated functional derivative with respect to this
interaction to satisfy
\begin{equation}
  \label{vert00}
  \frac{\delta V_{123}}{\delta V_{456}}=\delta_{14}\delta_{52}\delta_{36}.
\end{equation}
By introducing the graphical rule
\begin{equation}
\label{vert01}
\dvertex{}{2}{1}{3}\quad\equiv\quad\frac{\delta}{\delta V_{123}}  ,
\end{equation}
the definition of the functional derivative
(\ref{vert00}) can be expressed as
\begin{equation}
\label{vert02}
\dvertex{}{5}{4}{6}
\parbox{15mm}{\centerline{
\begin{fmfgraph*}(5,4.33)
\setval
\fmfforce{1w,0h}{v1}
\fmfforce{0w,0h}{v2}
\fmfforce{0.5w,1h}{v3}
\fmfforce{0.5w,0.2886h}{vm}
\fmf{fermion}{v1,vm,v2}
\fmf{boson}{v3,vm}
\fmfv{decor.size=0,label={\footnotesize 2},l.dist=0.5mm}{v1}
\fmfv{decor.size=0,label={\footnotesize 1},l.dist=0.5mm}{v2}
\fmfv{decor.size=0,label={\footnotesize 3},l.dist=0.5mm}{v3}
\fmfdot{vm}
\end{fmfgraph*}
}}\quad = \quad
\parbox{15mm}{\centerline{
\begin{fmfgraph*}(12,10.392)
\setval
\fmfforce{1w,0h}{v1}
\fmfforce{0.7w,0.2h}{v1b}
\fmfforce{0.85w,0.1h}{v1c}
\fmfforce{0w,0h}{v2}
\fmfforce{0.3w,0.2h}{v2b}
\fmfforce{0.15w,0.1h}{v2c}
\fmfforce{0.5w,1h}{v3}
\fmfforce{0.5w,0.639h}{v3b}
\fmfforce{0.5w,0.82h}{v3c}
\fmf{boson}{v3,v3c,v3b}
\fmf{fermion}{v1,v1c,v1b}
\fmf{fermion}{v2b,v2c,v2}
\fmfv{decor.size=0,label={\footnotesize 2},l.dist=0.5mm}{v1}
\fmfv{decor.size=0,label={\footnotesize 1},l.dist=0.5mm}{v2}
\fmfv{decor.size=0,label={\footnotesize 3},l.dist=0.5mm}{v3}
\fmfv{decor.size=0,label={\footnotesize 5},l.dist=0.5mm,l.angle=120}{v1b}
\fmfv{decor.size=0,label={\footnotesize 4},l.dist=0.5mm,l.angle=60}{v2b}
\fmfv{decor.size=0,label={\footnotesize 6},l.dist=0.5mm,l.angle=-90}{v3b}
\fmfv{decor.shape=circle,decor.filled=empty,decor.size=0.6mm}{v1c}
\fmfv{decor.shape=circle,decor.filled=empty,decor.size=0.6mm}{v2c}
\fmfv{decor.shape=circle,decor.filled=empty,decor.size=0.6mm}{v3c}
\end{fmfgraph*}
}},
\end{equation}
where the right-hand side represents a product of $\delta$-functions as 
defined in Eqs.~(\ref{rr19d}) and (\ref{rr30a}).
\subsection{Vacuum Energy as Generating Functional}
With the above-introduced diagrammatic operations,
the vacuum energy $W[S^{-1},D^{-1},V]$ constitutes a generating functional
for all correlation functions. Its evaluation
proceeds by expanding  
the exponential in the partition function
(\ref{gt18}) in powers of the coupling constant $e$, leading to the 
Taylor series
\begin{eqnarray}
\label{vac00}
Z = \sum_{p=0}^{\infty} \frac{e^{2p}}{(2p)!} \, \oint\meas 
\left( \int_{1\cdots 6} V_{123} V_{456}
\bar{\psi}_1 \psi_2 A_3 \bar{\psi}_4 \psi_5 A_6 \right)^p
e^{- {\cal A}^{(0)} [ \bar{\psi}, \psi, A;S^{-1},D^{-1}]}.
\end{eqnarray}
The products of pairs of fields $\bar{\psi}_1 \psi_2$ 
and $A_3 A_6$ can be substituted by a functional derivative with 
respect to $S^{-1}$ and $D^{-1}$, leading to the perturbation
expansion
\begin{eqnarray}
\label{vac01}
Z \equiv\sum\limits_{p=0}^\infty e^{2p}Z^{(p)}= 
\sum_{p=0}^{\infty} \frac{(-2 e^2)^p}{(2p)!} \, 
\left( \,\int_{1\cdots 6} V_{123} V_{456} \, \frac{\delta^3}{\delta 
S^{-1}_{12}
\delta S^{-1}_{45} \delta D^{-1}_{36}} \right)^p \, Z^{(0)}.
\end{eqnarray}
Note the two advantages of this expansion over the conventional one
in terms of currents coupled linearly 
to the fields. First, it contains only half as many 
functional derivatives.
Second, it does not contain derivatives with respect
to Grassmann variables.

Inserting for $Z^{(0)}$ the free vacuum
functional (\ref{gt19b}), we obtain for 
the first-order term $Z^{(1)}$
\begin{equation}
  \label{vac01b}
  Z^{(1)}=\frac{1}{2!}\int_{1\cdots 6}V_{123}V_{456}
(-2)\,\frac{\delta^3}{\delta D^{-1}_{36}\delta S^{-1}_{12}\delta S^{-1}_{45}}
\,Z^{(0)}.
\end{equation}
Since
\begin{equation}
  \label{vac01c}
  Z=Z^{(0)}+ e^2 Z^{(1)}+\ldots=\exp\left\{W^{(0)}+
 e^2 W^{(1)}+\ldots \right\},
\end{equation}
this corresponds to a first-order correction $W^{(1)}$ to the vacuum 
energy $W^{(0)}$:
\begin{equation}
  \label{vac04}
  W^{(1)}=\frac{1}{2!}\int_{1\ldots 6}V_{123}V_{456}(-2)
\frac{\delta W_{A}^{(0)}}
{\delta D^{-1}_{36}}\left(\frac{\delta^2 W_{\psi}^{(0)}}{\delta S^{-1}_{12}
\delta S^{-1}_{45}}+\frac{\delta W_{\psi}^{(0)}}{\delta S^{-1}_{12}}
\frac{\delta W_{\psi}^{(0)}}{\delta S^{-1}_{45}}\right).
\end{equation}
Expressing the derivatives with respect to the kernels by the 
corresponding propagators via Eqs.~(\ref{rr09}), (\ref{rr21}), and taking
into account (\ref{rr26}), $W^{(1)}$ becomes
\begin{equation}
  \label{vac05}
  W^{(1)}=\frac{1}{2}\int_{1\ldots 6}V_{123}V_{456}\,D_{36}
\left(S_{21}S_{54}-S_{24}S_{51} \right).
\end{equation}
According to the Feynman rules (\ref{vac07})--(\ref{vac09}), 
this is represented by the diagrams
\begin{eqnarray}
\label{vac06}
W^{(1)} = \,\frac{1}{2} \,
\parbox{19mm}{\centerline{
\begin{fmfgraph}(15,7)
\setval
\fmfforce{0.33w,0.5h}{v1}
\fmfforce{0.66w,0.5h}{v2}
\fmf{boson}{v1,v2}
\fmfi{fermion}{reverse fullcircle scaled 0.33w shifted (0.165w,0.5h)}
\fmfi{fermion}{fullcircle rotated 180 scaled 0.33w shifted (0.825w,0.5h)}
\fmfdot{v1,v2}
\end{fmfgraph}
}}-\, \frac{1}{2} \, 
\parbox{10mm}{\centerline{
\begin{fmfgraph}(7,5)
\setval
\fmfleft{v1}
\fmfright{v2}
\fmf{boson}{v1,v2}
\fmf{fermion,left=0.7}{v2,v1}
\fmf{fermion,left=0.7}{v1,v2}
\fmfdot{v1,v2}
\end{fmfgraph}
}} .
\end{eqnarray}
Note that each closed electron loop causes a
factor $-1$.
\section{Graphical Recursion Relation For Connected Vacuum Diagrams}
\label{recrel}
In this section, we derive a functional differential equation for the 
vacuum functional $W[S^{-1},D^{-1},V]$ whose solution leads to a graphical
recursion relation for all connected vacuum diagrams.
\subsection{Functional Differential Equation for $W = \ln Z$}
The functional differential equation for the vacuum functional
$W[S^{-1},D^{-1},V]$ is derived from the following functional integral
identity
\begin{equation}
\label{vac11} 
\oint \meas\, \frac{\delta}{\delta \bar{\psi}_1}\left\{\bar{\psi}_2\,
{\rm e}^{-{\cal A}[\bar{\psi},\psi,A;S^{-1},D^{-1},V]}\right\}=0
\end{equation}
with the action (\ref{gt18b}). This identity is the functional 
generalization of the trivial integral identity 
$\int_{-\infty}^{+\infty} dx\,f'(x)=0$ for
functions $f(x)$ which vanish at infinity. Nontrivial consequences of
Eq.~(\ref{vac11}) are obtained by
performing the functional derivative in the integrand which yields
\begin{eqnarray}
\label{vac11b}
\oint \meas  \left\{ \delta_{12} +
\int_3 \bar{\psi}_2 S^{-1}_{13} \psi_3 - e \int_{34} V_{134} \bar{\psi}_2
\psi_3 A_4 \right\} \,  {\rm e}^{- {\cal A} [ \bar{\psi}, 
\psi, A;S^{-1},D^{-1},V]} = 0.
\end{eqnarray}
Substituting the field product $\bar{\psi}_2 \psi_3$ by functional 
derivatives with respect to the electron kernel $S^{-1}_{23}$, this equation 
can be expressed in terms of the partition function (\ref{gt18}):
\begin{eqnarray}
\label{vac12}
Z\delta_{12} - \int_3 S^{-1}_{13} \, \frac{\delta Z}{\delta S^{-1}_{23}}
+ e \int_{34} V_{134} \, \frac{\delta}{ \delta S^{-1}_{23}}
[ \langle \hat{A}_4 \rangle Z] = 0 .
\end{eqnarray}
To bring this functional differential equation into a more convenient 
form, we calculate explicitly 
the term containing the expectation of the field $A$. 
This is done starting from
the integral identity
\begin{equation}
  \label{vac13}
  \oint\meas\,\frac{\delta}{\delta A_1}\,{\rm e}^{-{\cal A}[\bar{\psi},
\psi,A;S^{-1},D^{-1},V]}=0.
\end{equation}
Note this identity is not endangered by the gauge freedom in the electromagnetic
vector potential $A_\mu$ due to the presence of a gauge fixing term in the 
action~(\ref{gt01}). This ensures that the exponential vanishes at the boundary
of all $A$ field directions~\cite{comment1}.

After differentiating the action in the exponential of Eq.~(\ref{vac13}), 
we find the expectation of the photon field
\begin{equation}
  \label{vac13b}
  \int_1 \mean{\hat{A}_1}D^{-1}_{12}=- e \int_{34}V_{342}\mean{\psi_4
\bar{\psi}_3}.
\end{equation}
Multiplying this with $\int_2 D_{25}$, we yield
\begin{equation}
  \label{vac14}
  \mean{\hat{A}_5}=- e \int_{234} V_{342} D_{25}
\frac{\delta W}{\delta S^{-1}_{34}},
\end{equation}
where we have used $Z={\rm e}^W$. Inserting this into 
Eq.~(\ref{vac12}), we obtain 
\begin{equation}
  \label{vac15}
  \delta_{12}-\int_{3}S^{-1}_{13}\frac{\delta W}{\delta S^{-1}_{23}}=
e^2\int_{3\cdots 7} V_{134}V_{567}D_{47}\left\{
\frac{\delta^2W}{\delta S^{-1}_{23}\delta S^{-1}_{56}}+
\frac{\delta W}{\delta S^{-1}_{23}}\frac{\delta W}{\delta S^{-1}_{56}} 
\right\}.
\end{equation}
Setting $x_1=x_2$ and 
performing the integration over $x_1$, this leads to the nonlinear
functional differential equation for the vacuum functional $W$
\begin{equation}
  \label{vac15b}
  \int_1\delta_{11}-\int_{12}S^{-1}_{12}\frac{\delta W}{\delta S^{-1}_{12}}=
e^2\int_{1\cdots 6} V_{123}V_{456}D_{36}\left\{
\frac{\delta^2W}{\delta S^{-1}_{12}\delta S^{-1}_{45}}+
\frac{\delta W}{\delta S^{-1}_{12}}\frac{\delta W}{\delta S^{-1}_{45}} 
\right\},
\end{equation}
which will 
form the basis for deriving the desired
recursion relation for the vacuum diagrams. 
The first term on the left-hand side of Eq.~(\ref{vac15b}) is infinite, 
but in the next section we will show that this cancels 
against an infinity in the second term.
\subsection{Recursion Relation}
Equation (\ref{vac15b}) contains 
functional derivatives with 
respect to the electron kernel $S^{-1}$ which are equivalent
to cutting lines in the vacuum diagrams.
For practical purposes it will be 
more convenient to work with
derivatives with respect to the 
propagators $S$ which remove 
electron lines.
The second term on the left-hand side of Eq.~(\ref{vac15b}) contains 
the operation $-\int_{12}S^{-1}_{12}\delta/\delta S^{-1}_{12}$, 
which we convert into the differential operator
\begin{equation}
  \label{vac17}
  \hat{N}_{\rm F}=\int_{12}S_{12}\frac{\delta}{\delta S_{12}}
\end{equation}
with the help of (\ref{rr31}).
This operator has a simple graphical interpretation.
The derivative $\delta/\delta S_{12}$ removes an electron line from a 
Feynman diagram, and the factor $S_{12}$ restores it. 
This operation is familiar from
the number operator in 
second quantization. The operator $\hat{N}_{\rm F}$ counts the number 
of electron lines in a
Feynman diagram $G$:
\begin{equation}
  \label{vac18}
  \hat{N}_{\rm F}G=N_{\rm F}G.
\end{equation}
When applied to the vacuum diagrams $W^{(p)}$ of order $p\ge 1$, 
this operator gives
\begin{equation}
  \label{vac19}
  \hat{N}_{\rm F}W^{(p)}=2p\,W^{(p)},\quad p\ge 1,
\end{equation}
since the number of electron lines in a vacuum diagram without
external sources in quantum 
electrodynamics is equal to the number of vertices.
The restriction in Eq.~(\ref{vac19}) to $p\ge 1$ is necessary due to a 
special role of the vacuum diagram.
Take, for example, the electron vacuum diagram of the free 
theory (\ref{gt19d}). By applying the
operator $\hat{N}_{\rm F}$, we 
obtain with (\ref{rr21})
\begin{equation}
  \label{vac20}
  \hat{N}_{\rm F}W_{\psi}^{(0)}=-\int_{12}S^{-1}_{12}S_{21}=
-\int_1\delta_{11},
\end{equation}
which is a divergent trace integral precisely canceling
the infinite first term in Eq.~(\ref{vac15b}). 

Separating out $W^{(0)}$ in the expansion 
(\ref{gt19f}) of the vacuum functional, 
the left-hand side of the functional differential equation (\ref{vac15b}) has
the expansion
\begin{equation}
  \label{vac22}
  \int_1\delta_{11}+\hat{N}_{\rm F}W=\sum\limits_{p=1}^\infty\,2p \,
e^{2p} \, W^{(p)}=
\sum\limits_{p=0}^\infty\,2(p+1) \, e^{2(p+1)}\,W^{(p+1)}.
\end{equation}
On the right hand side of Eq.~(\ref{vac15b}), 
we express the first and second 
functional derivatives with respect to the kernel
$S^{-1}$ in terms of functional derivatives with respect to the propagator
$S$ by using Eq.~(\ref{rr31}) and
\begin{equation}
\frac{\delta^2}{\delta S^{-1}_{12} \delta S^{-1}_{34} } = \int_{5678}
\, S_{51} S_{26} S_{73} S_{48} \, 
\frac{\delta^2}{\delta S_{56} \delta S_{78} }
+ \int_{56} \left[ S_{53} S_{41} S_{26} + S_{23} S_{46} S_{51} \right] \,
\frac{\delta}{\delta S_{56}} \, .
\end{equation}
Inserting here the expansion (\ref{gt19f}) and comparing equal powers
in $e$ with those in Eq.~(\ref{vac22}), we obtain  
the following recursion formula for the expansion coefficients of the 
vacuum functional 
\begin{eqnarray}
  \label{vac23}
  W^{(p+1)}=\frac{1}{2(p+1)}\Bigg\{&&\int_{1\ldots 10}V_{123}V_{456}D_{36}
S_{71}S_{28}S_{94}S_{5\,10}\frac{\delta^2 W^{(p)}}{\delta S_{78}\delta 
S_{9\,10}}\nonumber\\
&&+2\int_{1\ldots 8}V_{123}V_{456}D_{36}\left(S_{51}S_{28}S_{74}-S_{71}
S_{28}S_{54}\right)\frac{\delta W^{(p)}}{\delta S_{78}}\nonumber\\
&&+\sum\limits_{q=1}^{p-1}\,\int_{1\ldots 10}V_{123}V_{456}D_{36}S_{71}
S_{28}S_{94}S_{5\,10}\frac{\delta W^{(q)}}{\delta S_{78}}
\frac{\delta W^{(p-q)}}{\delta S_{9\,10}}\Bigg\},\quad p\ge 1
\end{eqnarray}
and the initial value (\ref{vac05}).
This equation enables us to derive the connected vacuum diagrams 
systematically to any desired order from the diagrams of the previous 
orders, as will now be shown.
\subsection{Graphical Solution}
With the help of the Feynman rules (\ref{vac07})--(\ref{vac09}),
the functional recursion relation (\ref{vac23}) can
be written diagrammatically as follows
\begin{eqnarray}
\label{vac26}
W^{(p+1)}\equiv\frac{1}{2(p+1)}\,&&\Bigg\{
\parbox{14mm}{\begin{center}
\begin{fmfgraph*}(8,9)
\setval
\fmfstraight
\fmfforce{0w,0.835h}{i1}
\fmfforce{0w,0.165h}{i2}
\fmfforce{1w,1h}{o1}
\fmfforce{1w,0.66h}{o2}
\fmfforce{1w,0.33h}{o3}
\fmfforce{1w,0h}{o4}
\fmf{boson}{i1,i2}
\fmf{fermion,left=0.1}{i1,o1}
\fmf{fermion,left=0.1}{o2,i1}
\fmf{fermion,left=0.1}{i2,o3}
\fmf{fermion,left=0.1}{o4,i2}
\fmfdot{i1,i2}
\fmfv{decor.size=0, label=${\scs 1}$, l.dist=1mm, l.angle=0}{o1}
\fmfv{decor.size=0, label=${\scs 2}$, l.dist=1mm, l.angle=0}{o2}
\fmfv{decor.size=0, label=${\scs 3}$, l.dist=1mm, l.angle=0}{o3}
\fmfv{decor.size=0, label=${\scs 4}$, l.dist=1mm, l.angle=0}{o4}
\end{fmfgraph*}
\end{center}}
\quad \ddfermi{W^{(p)}}\; +\;2\;\Bigg[
\parbox{13mm}{\begin{center}
\begin{fmfgraph*}(8,6)
\setval
\fmfstraight
\fmfforce{0.3w,1h}{i1}
\fmfforce{0.3w,0h}{i2}
\fmfforce{1w,1h}{o1}
\fmfforce{1w,0h}{o2}
\fmf{fermion}{i2,i1}
\fmf{boson,right=0.7}{i1,i2}
\fmf{fermion}{i1,o1}
\fmf{fermion}{o2,i2}
\fmfdot{i1,i2}
\fmfv{decor.size=0, label=${\scs 1}$, l.dist=1mm, l.angle=0}{o1}
\fmfv{decor.size=0, label=${\scs 2}$, l.dist=1mm, l.angle=0}{o2}
\end{fmfgraph*}
\end{center}}
\quad-\;
\parbox{17mm}{\begin{center}
\begin{fmfgraph*}(12,6)
\setval
\fmfstraight
\fmfforce{0.33w,0.5h}{v1}
\fmfforce{0.66w,0.5h}{v2}
\fmfforce{1w,1h}{o1}
\fmfforce{1w,0h}{o2}
\fmf{boson}{v1,v2}
\fmf{fermion}{v2,o1}
\fmf{fermion}{o2,v2}
\fmfi{fermion}{reverse fullcircle scaled 0.33w shifted (0.165w,0.5h)}
\fmfdot{v1,v2}
\fmfv{decor.size=0, label=${\scs 1}$, l.dist=1mm, l.angle=0}{o1}
\fmfv{decor.size=0, label=${\scs 2}$, l.dist=1mm, l.angle=0}{o2}
\end{fmfgraph*}
\end{center}}
\quad\Bigg] \dfermi{W^{(p)}}{1}{2}\nonumber\\
&&+\;\sum\limits_{q=1}^{p-1}\quad\dfermi{W^{(p-q)}}{1}{2}\quad
\parbox{17mm}{\begin{center}
\begin{fmfgraph*}(12,6)
\setval
\fmfstraight
\fmfleft{i2,i1}
\fmfright{o2,o1}
\fmf{fermion}{v1,i1}
\fmf{fermion}{i2,v1}
\fmf{boson}{v1,v2}
\fmf{fermion}{v2,o1}
\fmf{fermion}{o2,v2}
\fmfdot{v1,v2}
\fmfv{decor.size=0, label=${\scs 1}$, l.dist=1mm, l.angle=-180}{i1}
\fmfv{decor.size=0, label=${\scs 2}$, l.dist=1mm, l.angle=-180}{i2}
\fmfv{decor.size=0, label=${\scs 3}$, l.dist=1mm, l.angle=0}{o1}
\fmfv{decor.size=0, label=${\scs 4}$, l.dist=1mm, l.angle=0}{o2}
\end{fmfgraph*}
\end{center}}
\quad\dfermi{W^{(q)}}{3}{4}\;\Bigg\},\qquad p\ge 1 
\end{eqnarray}
and the first-order result is given by Eq.~(\ref{vac06}). 
The right-hand side 
contains four graphical operations. The first three are linear
and involve one or two electron line amputations of the previous perturbative
order. The fourth operation is nonlinear and mixes two different
electron line amputations of lower orders.
To demonstrate the working of this formula, we calculate 
the connected
vacuum diagrams in second and third order. 
We start with the amputation of one or two electron lines in first order
(\ref{vac06}):
\begin{eqnarray}
\label{vac27}
\dfermi{W^{(1)}}{1}{2}\,\,\,\,\equiv \quad
\parbox{17mm}{\begin{center}
\begin{fmfgraph*}(12,6)
\setval
\fmfstraight
\fmfforce{0.33w,0.5h}{v1}
\fmfforce{0.66w,0.5h}{v2}
\fmfforce{0w,1h}{i1}
\fmfforce{0w,0h}{i2}
\fmf{boson}{v1,v2}
\fmf{fermion}{i1,v1}
\fmf{fermion}{v1,i2}
\fmfi{fermion}{fullcircle rotated 180 scaled 0.33w shifted (0.825w,0.5h)}
\fmfdot{v1,v2}
\fmfv{decor.size=0, label=${\scs 1}$, l.dist=1mm, l.angle=-180}{i1}
\fmfv{decor.size=0, label=${\scs 2}$, l.dist=1mm, l.angle=-180}{i2}
\end{fmfgraph*}
\end{center}}
\;-\quad
\parbox{13mm}{\begin{center}
\begin{fmfgraph*}(8,6)
\setval
\fmfstraight
\fmfforce{0.7w,1h}{o1}
\fmfforce{0.7w,0h}{o2}
\fmfforce{0w,1h}{i1}
\fmfforce{0w,0h}{i2}
\fmf{fermion}{o1,o2}
\fmf{boson,left=0.7}{o1,o2}
\fmf{fermion}{i1,o1}
\fmf{fermion}{o2,i2}
\fmfdot{o1,o2}
\fmfv{decor.size=0, label=${\scs 1}$, l.dist=1mm, l.angle=-180}{i1}
\fmfv{decor.size=0, label=${\scs 2}$, l.dist=1mm, l.angle=-180}{i2}
\end{fmfgraph*}
\end{center}}, \quad \quad
\ddfermi{W^{(1)}}\quad\equiv \quad
\parbox{14mm}{\begin{center}
\begin{fmfgraph*}(8,9)
\setval
\fmfstraight
\fmfforce{1w,0.835h}{o1}
\fmfforce{1w,0.165h}{o2}
\fmfforce{0w,1h}{i1}
\fmfforce{0w,0.66h}{i2}
\fmfforce{0w,0.33h}{i3}
\fmfforce{0w,0h}{i4}
\fmf{boson}{o1,o2}
\fmf{fermion,left=0.1}{i1,o1}
\fmf{fermion,left=0.1}{o1,i2}
\fmf{fermion,left=0.1}{i3,o2}
\fmf{fermion,left=0.1}{o2,i4}
\fmfdot{o1,o2}
\fmfv{decor.size=0, label=${\scs 1}$, l.dist=1mm, l.angle=-180}{i1}
\fmfv{decor.size=0, label=${\scs 2}$, l.dist=1mm, l.angle=-180}{i2}
\fmfv{decor.size=0, label=${\scs 3}$, l.dist=1mm, l.angle=-180}{i3}
\fmfv{decor.size=0, label=${\scs 4}$, l.dist=1mm, l.angle=-180}{i4}
\end{fmfgraph*}
\end{center}}
-\quad 
\parbox{14mm}{\begin{center}
\begin{fmfgraph*}(8,9)
\setval
\fmfstraight
\fmfforce{1w,0.835h}{o1}
\fmfforce{1w,0.165h}{o2}
\fmfforce{0w,1h}{i1}
\fmfforce{0w,0.66h}{i2}
\fmfforce{0w,0.33h}{i3}
\fmfforce{0w,0h}{i4}
\fmf{boson}{o1,o2}
\fmf{fermion,left=0.1}{i1,o1}
\fmf{fermion,left=0.1}{o1,i2}
\fmf{fermion,left=0.1}{i3,o2}
\fmf{fermion,left=0.1}{o2,i4}
\fmfdot{o1,o2}
\fmfv{decor.size=0, label=${\scs 1}$, l.dist=1mm, l.angle=-180}{i1}
\fmfv{decor.size=0, label=${\scs 4}$, l.dist=1mm, l.angle=-180}{i2}
\fmfv{decor.size=0, label=${\scs 3}$, l.dist=1mm, l.angle=-180}{i3}
\fmfv{decor.size=0, label=${\scs 2}$, l.dist=1mm, l.angle=-180}{i4}
\end{fmfgraph*}
\end{center}}.
\end{eqnarray}
Inserting (\ref{vac27}) into (\ref{vac26}), where we 
have to take care of connecting 
only legs with the same label, we find the second-order correction of the
vacuum functional $W$:
\end{fmffile}
\begin{fmffile}{diag2bkpB}
\begin{eqnarray}
\label{vac29}
W^{(2)}\equiv \frac{1}{4}
\parbox{18mm}{\begin{center}
\begin{fmfgraph}(8,5)
\setval
\fmfleft{i2,i1}
\fmfright{o2,o1}
\fmf{fermion,left=1}{i1,i2,i1}
\fmf{boson}{i1,o1}
\fmf{boson}{i2,o2}
\fmf{fermion,left=1}{o1,o2,o1}
\fmfdot{i1,i2,o1,o2}
\end{fmfgraph}
\end{center}}
+
\parbox{18mm}{\begin{center}
\begin{fmfgraph}(13,5)
\setval
\fmfforce{0.385w,0.5h}{v1}
\fmfforce{0.615w,0.5h}{v2}
\fmfforce{0.192w,1h}{v3}
\fmfforce{0.192w,0h}{v4}
\fmf{fermion,left=0.5}{v3,v1,v4}
\fmf{fermion,left=1}{v4,v3}
\fmf{boson}{v3,v4}
\fmf{boson}{v1,v2}
\fmfi{fermion}{fullcircle rotated 180 scaled 0.384w shifted (0.808w,0.5h)}
\fmfdot{v1,v2,v3,v4}
\end{fmfgraph}
\end{center}}
-\,\frac{1}{2}
\parbox{26mm}{\begin{center}
\begin{fmfgraph}(21,5)
\setval
\fmfforce{0.238w,0.5h}{v1}
\fmfforce{0.381w,0.5h}{v2}
\fmfforce{0.619w,0.5h}{v3}
\fmfforce{0.762w,0.5h}{v4}
\fmf{fermion,left=1}{v2,v3,v2}
\fmf{boson}{v1,v2}
\fmf{boson}{v3,v4}
\fmfi{fermion}{reverse fullcircle scaled 0.238w shifted (0.119w,0.5h)}
\fmfi{fermion}{fullcircle rotated 180 scaled 0.238w shifted (0.881w,0.5h)}
\fmfdot{v1,v2,v3,v4}
\end{fmfgraph}
\end{center}}
-\,\frac{1}{4}
\parbox{14mm}{\begin{center}
\begin{fmfgraph}(5,5)
\setval
\fmfleft{i2,i1}
\fmfright{o2,o1}
\fmf{fermion,left=0.5}{i1,o1,o2,i2,i1}
\fmf{boson}{i1,o2}
\fmf{boson}{i2,o1}
\fmfdot{i1,i2,o1,o2}
\end{fmfgraph}
\end{center}}
-\,\frac{1}{2}
\parbox{14mm}{\begin{center}
\begin{fmfgraph}(5,5)
\setval
\fmfleft{i2,i1}
\fmfright{o2,o1}
\fmf{fermion,left=0.5}{i1,o1,o2,i2,i1}
\fmf{boson,left=0.4}{i1,i2}
\fmf{boson,right=0.4}{o1,o2}
\fmfdot{i1,i2,o1,o2}
\end{fmfgraph}
\end{center}}. 
\end{eqnarray}
The calculation of the third-order correction $W^{(3)}$ 
leads to the following 20 diagrams:
\begin{eqnarray}
\label{vac29b}
W^{(3)}&\equiv& \,\frac{1}{2}
\parbox{18mm}{\begin{center}
\begin{fmfgraph}(13,5)
\setval
\fmfforce{0w,0.5h}{i1}
\fmfforce{0.192w,1h}{i2}
\fmfforce{0.385w,0.5h}{i3}
\fmfforce{0.192w,0h}{i4}
\fmfforce{0.808w,1h}{o1}
\fmfforce{0.808w,0h}{o2}
\fmf{fermion,left=0.5}{i1,i2,i3,i4,i1}
\fmf{boson}{i1,i3}
\fmf{boson}{i2,o1}
\fmf{boson}{i4,o2}
\fmf{fermion,left=1}{o1,o2,o1}
\fmfdot{i1,i2,i3,i4,o1,o2}
\end{fmfgraph}
\end{center}}
+\,\frac{1}{6}
\parbox{18mm}{\begin{center}
\begin{fmfgraph}(13,5)
\setval
\fmfforce{0.192w,1h}{i1}
\fmfforce{0.385w,0.5h}{i2}
\fmfforce{0.192w,0h}{i3}
\fmfforce{0.808w,1h}{o1}
\fmfforce{0.808w,0h}{o2}
\fmfforce{0.615w,0.5h}{o3}
\fmf{fermion,left=0.5}{i1,i2,i3}
\fmf{fermion,left=1}{i3,i1}
\fmf{boson}{i1,o1}
\fmf{boson}{i2,o3}
\fmf{boson}{i3,o2}
\fmf{fermion,left=1}{o1,o2}
\fmf{fermion,left=0.5}{o2,o3,o1}
\fmfdot{i1,i2,i3,o1,o2,o3}
\end{fmfgraph}
\end{center}}
+\,\frac{1}{6}
\parbox{18mm}{\begin{center}
\begin{fmfgraph}(13,5)
\setval
\fmfforce{0.192w,1h}{i1}
\fmfforce{0.385w,0.5h}{i2}
\fmfforce{0.192w,0h}{i3}
\fmfforce{0.808w,1h}{o1}
\fmfforce{0.808w,0h}{o2}
\fmfforce{0.615w,0.5h}{o3}
\fmf{fermion,left=0.5}{i1,i2,i3}
\fmf{fermion,left=1}{i3,i1}
\fmf{boson}{i1,o1}
\fmf{boson}{i2,o3}
\fmf{boson}{i3,o2}
\fmf{fermion,right=1}{o2,o1}
\fmf{fermion,right=0.5}{o1,o3,o2}
\fmfdot{i1,i2,i3,o1,o2,o3}
\end{fmfgraph}
\end{center}}
+
\parbox{18mm}{\begin{center}
\begin{fmfgraph}(13,5)
\setval
\fmfforce{0.192w,1h}{i1}
\fmfforce{0.192w,0h}{i2}
\fmfforce{0.026w,0.75h}{i3}
\fmfforce{0.026w,0.25h}{i4}
\fmfforce{0.808w,1h}{o1}
\fmfforce{0.808w,0h}{o2}
\fmf{fermion,left=1}{i1,i2}
\fmf{fermion,left=1/3}{i2,i4,i3,i1}
\fmf{boson,left=0.8}{i3,i4}
\fmf{boson}{i1,o1}
\fmf{boson}{i2,o2}
\fmf{fermion,left=1}{o1,o2,o1}
\fmfdot{i1,i2,i3,i4,o1,o2}
\end{fmfgraph}
\end{center}}
-\,\frac{1}{6}
\parbox{20mm}{\begin{center}
\begin{fmfgraph}(10,5)
\setval
\fmfforce{0w,1h}{i1}
\fmfforce{0w,0h}{i2}
\fmfforce{0.33w,0h}{v1}
\fmfforce{0.66w,0h}{v2}
\fmfforce{1w,1h}{o1}
\fmfforce{1w,0h}{o2}
\fmf{fermion,left=1}{i1,i2,i1}
\fmf{fermion,left=1}{o1,o2,o1}
\fmf{boson}{i1,o1}
\fmf{boson}{i2,v1}
\fmf{fermion,left=1}{v1,v2,v1}
\fmf{boson}{v2,o2}
\fmfdot{i1,i2,v1,v2,o1,o2}
\end{fmfgraph}
\end{center}}
+\,\frac{1}{2}
\parbox{18mm}{\begin{center}
\begin{fmfgraph}(13,5)
\setval
\fmfforce{0.192w,1h}{i1}
\fmfforce{0.385w,0.5h}{i2}
\fmfforce{0.192w,0h}{i3}
\fmfforce{0.808w,1h}{o1}
\fmfforce{0.808w,0h}{o2}
\fmfforce{0.615w,0.5h}{o3}
\fmf{fermion,left=0.5}{i1,i2,i3}
\fmf{fermion,left=1}{i3,i1}
\fmf{boson}{i1,i3}
\fmf{boson}{i2,o3}
\fmf{boson}{o1,o2}
\fmf{fermion,right=1}{o2,o1}
\fmf{fermion,right=0.5}{o1,o3,o2}
\fmfdot{i1,i2,i3,o1,o2,o3}
\end{fmfgraph}
\end{center}}
\nonumber\\
&&+
\parbox{17mm}{\begin{center}
\begin{fmfgraph}(13,5)
\setval
\fmfforce{0.252w,0.976h}{i2}
\fmfforce{0.037w,0.794h}{i3}
\fmfforce{0.037w,0.206h}{i4}
\fmfforce{0.252w,0.024h}{i5}
\fmfforce{0.385w,0.5h}{i1}
\fmfforce{0.615w,0.5h}{o1}
\fmf{fermion,right=0.3}{i1,i2,i3,i4,i5,i1}
\fmf{boson}{i3,i5}
\fmf{boson}{i2,i4}
\fmf{boson}{i1,o1}
\fmfi{fermion}{fullcircle rotated 180 scaled 0.385w shifted (0.808w,0.5h)}
\fmfdot{i1,i2,i3,i4,i5,o1}
\end{fmfgraph}
\end{center}}
+
\parbox{17mm}{\begin{center}
\begin{fmfgraph}(13,5)
\setval
\fmfforce{0.252w,0.976h}{i2}
\fmfforce{0.037w,0.794h}{i3}
\fmfforce{0.037w,0.206h}{i4}
\fmfforce{0.252w,0.024h}{i5}
\fmfforce{0.385w,0.5h}{i1}
\fmfforce{0.615w,0.5h}{o1}
\fmf{fermion,right=0.3}{i1,i2,i3,i4,i5,i1}
\fmf{boson}{i2,i5}
\fmf{boson,left=0.7}{i3,i4}
\fmf{boson}{i1,o1}
\fmfi{fermion}{fullcircle rotated 180 scaled 0.385w shifted (0.808w,0.5h)}
\fmfdot{i1,i2,i3,i4,i5,o1}
\end{fmfgraph}
\end{center}}
+
\parbox{17mm}{\begin{center}
\begin{fmfgraph}(13,5)
\setval
\fmfforce{0.252w,0.976h}{i2}
\fmfforce{0.037w,0.794h}{i3}
\fmfforce{0.037w,0.206h}{i4}
\fmfforce{0.252w,0.024h}{i5}
\fmfforce{0.385w,0.5h}{i1}
\fmfforce{0.615w,0.5h}{o1}
\fmf{fermion,right=0.3}{i1,i2,i3,i4,i5,i1}
\fmf{boson,left=0.7}{i2,i3}
\fmf{boson,left=0.7}{i4,i5}
\fmf{boson}{i1,o1}
\fmfi{fermion}{fullcircle rotated 180 scaled 0.385w shifted (0.808w,0.5h)}
\fmfdot{i1,i2,i3,i4,i5,o1}
\end{fmfgraph}
\end{center}}
-
\parbox{25mm}{\begin{center}
\begin{fmfgraph}(21,5)
\setval
\fmfforce{0.238w,0.5h}{i1}
\fmfforce{0.381w,0.5h}{v1}
\fmfforce{0.5w,1h}{v2}
\fmfforce{0.5w,0h}{v3}
\fmfforce{0.881w,1h}{o1}
\fmfforce{0.881w,0h}{o2}
\fmfi{fermion}{reverse fullcircle scaled 0.238w shifted (0.119w,0.5h)}
\fmf{boson}{i1,v1}
\fmf{fermion,right=0.5}{v2,v1,v3}
\fmf{fermion,right=1}{v3,v2}
\fmf{boson}{v2,o1}
\fmf{boson}{v3,o2}
\fmf{fermion,right=1}{o1,o2,o1}
\fmfdot{i1,v1,v2,v3,o1,o2}
\end{fmfgraph}
\end{center}}
-\,\frac{1}{2}
\parbox{25mm}{\begin{center}
\begin{fmfgraph}(21,5)
\setval
\fmfforce{0.238w,0.5h}{v1}
\fmfforce{0.381w,0.5h}{v2}
\fmfforce{0.5w,1h}{v3}
\fmfforce{0.5w,0h}{v4}
\fmfforce{0.619w,0.5h}{v5}
\fmfforce{0.762w,0.5h}{v6}
\fmfi{fermion}{reverse fullcircle scaled 0.238w shifted (0.119w,0.5h)}
\fmfi{fermion}{fullcircle rotated 180 scaled 0.238w shifted (0.881w,0.5h)}
\fmf{boson}{v1,v2}
\fmf{fermion,left=0.5}{v2,v3,v5,v4,v2}
\fmf{boson}{v3,v4}
\fmf{boson}{v5,v6}
\fmfdot{v1,v2,v3,v4,v5,v6}
\end{fmfgraph}
\end{center}}
-
\parbox{25mm}{\begin{center}
\begin{fmfgraph}(21,5)
\setval
\fmfforce{0.238w,0.5h}{v1}
\fmfforce{0.381w,0.5h}{v2}
\fmfforce{0.5595w,0.933h}{v4}
\fmfforce{0.4405w,0.933h}{v3}
\fmfforce{0.619w,0.5h}{v5}
\fmfforce{0.762w,0.5h}{v6}
\fmfi{fermion}{reverse fullcircle scaled 0.238w shifted (0.119w,0.5h)}
\fmfi{fermion}{fullcircle rotated 180 scaled 0.238w shifted (0.881w,0.5h)}
\fmf{boson}{v1,v2}
\fmf{fermion,left=0.33}{v2,v3,v4,v5}
\fmf{fermion,left=1}{v5,v2}
\fmf{boson,right=0.8}{v3,v4}
\fmf{boson}{v5,v6}
\fmfdot{v1,v2,v3,v4,v5,v6}
\end{fmfgraph}
\end{center}}
\nonumber\\
&&-
\parbox{25mm}{\begin{center}
\begin{fmfgraph}(21,5)
\setval
\fmfforce{0.238w,0.5h}{v1}
\fmfforce{0.381w,0.5h}{v2}
\fmfforce{0.119w,1h}{v4}
\fmfforce{0.119w,0h}{v3}
\fmfforce{0.619w,0.5h}{v5}
\fmfforce{0.762w,0.5h}{v6}
\fmfi{fermion}{fullcircle rotated 180 scaled 0.238w shifted (0.881w,0.5h)}
\fmf{fermion,right=1}{v4,v3}
\fmf{fermion,right=0.45}{v3,v1,v4}
\fmf{boson}{v1,v2}
\fmf{fermion,left=1}{v2,v5,v2}
\fmf{boson}{v3,v4}
\fmf{boson}{v5,v6}
\fmfdot{v1,v2,v3,v4,v5,v6}
\end{fmfgraph}
\end{center}}
+\,\frac{1}{3}
\parbox{25mm}{\begin{center}
\begin{fmfgraph}(21,13)
\setval
\fmfforce{0.238w,0.192h}{v1}
\fmfforce{0.381w,0.192h}{v2}
\fmfforce{0.5w,0.384h}{v4}
\fmfforce{0.5w,0.615h}{v3}
\fmfforce{0.619w,0.192h}{v5}
\fmfforce{0.762w,0.192h}{v6}
\fmfi{fermion}{reverse fullcircle scaled 0.238w shifted (0.119w,0.192h)}
\fmfi{fermion}{fullcircle rotated 180 scaled 0.238w shifted (0.881w,0.192h)}
\fmfi{fermion}{fullcircle rotated 270 scaled 0.238w shifted (0.5w,0.808h)}
\fmf{boson}{v1,v2}
\fmf{fermion,left=0.5}{v2,v4,v5}
\fmf{fermion,left=1}{v5,v2}
\fmf{boson}{v3,v4}
\fmf{boson}{v5,v6}
\fmfdot{v1,v2,v3,v4,v5,v6}
\end{fmfgraph}
\end{center}}
+\,\frac{1}{2}
\parbox{32mm}{\begin{center}
\begin{fmfgraph}(29,5)
\setval
\fmfforce{0.172w,0.5h}{v1}
\fmfforce{0.276w,0.5h}{v2}
\fmfforce{0.448w,0.5h}{v3}
\fmfforce{0.552w,0.5h}{v4}
\fmfforce{0.724w,0.5h}{v5}
\fmfforce{0.828w,0.5h}{v6}
\fmfi{fermion}{reverse fullcircle scaled 0.172w shifted (0.086w,0.5h)}
\fmfi{fermion}{fullcircle rotated 180 scaled 0.172w shifted (0.914w,0.5h)}
\fmf{boson}{v1,v2}
\fmf{fermion,left=1}{v2,v3,v2}
\fmf{fermion,left=1}{v4,v5,v4}
\fmf{boson}{v3,v4}
\fmf{boson}{v5,v6}
\fmfdot{v1,v2,v3,v4,v5,v6}
\end{fmfgraph}
\end{center}}
-\,\frac{1}{2}
\parbox{11mm}{\begin{center}
\begin{fmfgraph}(7,7)
\setval
\fmfforce{0w,0.5h}{v1}
\fmfforce{0.25w,0.933h}{v2}
\fmfforce{0.75w,0.933h}{v3}
\fmfforce{1w,0.5h}{v4}
\fmfforce{0.75w,0.067h}{v5}
\fmfforce{0.25w,0.067h}{v6}
\fmf{fermion,left=0.3}{v1,v2,v3,v4,v5,v6,v1}
\fmf{boson}{v1,v4}
\fmf{boson}{v2,v6}
\fmf{boson}{v3,v5}
\fmfdot{v1,v2,v3,v4,v5,v6}
\end{fmfgraph}
\end{center}}
-\,\frac{1}{6}
\parbox{11mm}{\begin{center}
\begin{fmfgraph}(7,7)
\setval
\fmfforce{0w,0.5h}{v1}
\fmfforce{0.25w,0.933h}{v2}
\fmfforce{0.75w,0.933h}{v3}
\fmfforce{1w,0.5h}{v4}
\fmfforce{0.75w,0.067h}{v5}
\fmfforce{0.25w,0.067h}{v6}
\fmf{fermion,left=0.3}{v1,v2,v3,v4,v5,v6,v1}
\fmf{boson}{v1,v4}
\fmf{boson}{v2,v5}
\fmf{boson}{v3,v6}
\fmfdot{v1,v2,v3,v4,v5,v6}
\end{fmfgraph}
\end{center}}
-
\parbox{11mm}{\begin{center}
\begin{fmfgraph}(7,7)
\setval
\fmfforce{0w,0.5h}{v1}
\fmfforce{0.25w,0.933h}{v2}
\fmfforce{0.75w,0.933h}{v3}
\fmfforce{1w,0.5h}{v4}
\fmfforce{0.75w,0.067h}{v5}
\fmfforce{0.25w,0.067h}{v6}
\fmf{fermion,left=0.3}{v1,v2,v3,v4,v5,v6,v1}
\fmf{boson,right=0.7}{v2,v3}
\fmf{boson,right=0.2}{v4,v6}
\fmf{boson,right=0.2}{v5,v1}
\fmfdot{v1,v2,v3,v4,v5,v6}
\end{fmfgraph}
\end{center}}
\nonumber\\
&&-\,\frac{1}{2}
\parbox{11mm}{\begin{center}
\begin{fmfgraph}(7,7)
\setval
\fmfforce{0w,0.5h}{v1}
\fmfforce{0.25w,0.933h}{v2}
\fmfforce{0.75w,0.933h}{v3}
\fmfforce{1w,0.5h}{v4}
\fmfforce{0.75w,0.067h}{v5}
\fmfforce{0.25w,0.067h}{v6}
\fmf{fermion,left=0.3}{v1,v2,v3,v4,v5,v6,v1}
\fmf{boson,right=0.7}{v2,v3}
\fmf{boson}{v1,v4}
\fmf{boson,right=0.7}{v5,v6}
\fmfdot{v1,v2,v3,v4,v5,v6}
\end{fmfgraph}
\end{center}}
-\,\frac{1}{3}
\parbox{11mm}{\begin{center}
\begin{fmfgraph}(7,7)
\setval
\fmfforce{0w,0.5h}{v1}
\fmfforce{0.25w,0.933h}{v2}
\fmfforce{0.75w,0.933h}{v3}
\fmfforce{1w,0.5h}{v4}
\fmfforce{0.75w,0.067h}{v5}
\fmfforce{0.25w,0.067h}{v6}
\fmf{fermion,left=0.3}{v1,v2,v3,v4,v5,v6,v1}
\fmf{boson,right=0.7}{v2,v3}
\fmf{boson,right=0.7}{v4,v5}
\fmf{boson,right=0.7}{v6,v1}
\fmfdot{v1,v2,v3,v4,v5,v6}
\end{fmfgraph}
\end{center}}. 
\end{eqnarray}
From the vacuum diagrams (\ref{vac06}), (\ref{vac29}), and (\ref{vac29b}),
we observe a simple mnemonic rule 
for the weights of the connected vacuum diagrams in QED. At least up to four loops, 
each weight is
equal to the reciprocal number of electron lines, which, by cutting,
generate the same two-point diagrams.
The sign is given by $(-1)^L$, where $L$ denotes the number of electron
loops.
Note that the total weight, which is the sum over all weights of the 
vacuum diagrams in the order to be considered, vanishes in QED. 
The simplicity of the 
weights is a consequence of the Fermi statistics and the three-point 
form of interaction (\ref{gt09}). The weights of the vacuum diagrams
in other  
theories, like $\phi^4$-theory~\cite{phi4,Verena,Kleinert1}, 
follow more complicated rules.
\section{Scattering Between Electrons and Photons}
\label{npoint}
From the above vacuum diagrams, we obtain all even-point correlation 
functions
by cutting electron or photon lines. For the generation of the odd-point
functions we use the
functional derivative (\ref{vert02}) with respect to the
interaction function $V$ which removes a vertex from a 
diagram. 

As an illustration, we generate
the diagrams for the self interactions described by the propagators
(\ref{gt18d}) and (\ref{gt18e})
\begin{eqnarray}
\label{np00}
{}^\gamma G^{2}_{12} = \mean{\hat{A}_1\, \hat{A}_2}, \qquad
{}^eG^{2}_{12} = \mean{\hat{\psi}_1\, \hat{\bar{\psi}}_2}
\end{eqnarray}
and the four-point functions
\begin{eqnarray}
\label{np00b}
{}^{\gamma\gamma} G^{4}_{1234} = \mean{\hat{A}_1\, \hat{A}_2\, \hat{A}_3\, 
\hat{A}_4}, \quad 
{}^{ee}G^{4}_{1234} = \mean{\hat{\psi}_1\, \hat{\psi}_2\, 
\hat{\bar{\psi}}_3\, \hat{\bar{\psi}}_4}, \quad 
{}^{e\gamma}G^{4}_{1234} = \mean{\hat{\psi}_1\, \hat{A}_2\, \hat{A}_3\,
\hat{\bar{\psi}}_4}, 
\end{eqnarray} 
which represent the simplest scattering processes of the theory.
In addition, we give the perturbative expansion of the three-point
vertex function
\begin{equation}
  \label{np00c}
  G^{3}_{123}=\mean{\hat{\psi}_1\,\hat{\bar{\psi}}_2\,\hat{A}_3}.
\end{equation}
The following examples illustrate the simple weights $(-1)^L$ of diagrams contributing
to an $n$-point function with $n\ge 2$, with $L$ being the number of 
electron loops.
\subsection{Self Interactions}
Substituting the product of the photon fields $A_1 A_2$ in the functional
integral 
(\ref{gt18d}) by the photonic functional derivative 
$-2\delta/\delta D^{-1}_{12}$, 
the photonic two-point function of the interacting theory 
is given by 
\begin{equation}
  \label{np01}
  {}^\gamma G^{2}_{12}=-2\,\frac{\delta}{\delta D^{-1}_{12}}W[S^{-1},D^{-1},V].
\end{equation}
Applying the associated cutting rule (\ref{rr13}) 
to the vacuum diagrams (\ref{vac06}) and 
(\ref{vac29}) leads to the connected diagrams
\begin{eqnarray}
\label{np02}
{}^\gamma G^{2,c}_{12}&\equiv& 
\parbox{22mm}{\begin{center}
\begin{fmfgraph*}(9,3)
\setval
\fmfleft{v1}
\fmfright{v2}
\fmf{boson}{v1,v2}
\fmflabel{$\scs{1}$}{v1}
\fmflabel{$\scs{2}$}{v2}
\end{fmfgraph*}
\end{center}}-e^2
\parbox{18mm}{\begin{center}
\begin{fmfgraph}(15,5)
\setval
\fmfforce{0w,0.5h}{v1}
\fmfforce{1/3w,0.5h}{v2}
\fmfforce{2/3w,0.5h}{v3}
\fmfforce{1w,0.5h}{v4}
\fmf{boson}{v1,v2}
\fmf{fermion,right}{v2,v3,v2}
\fmf{boson}{v3,v4}
\fmfdot{v2,v3}
\end{fmfgraph}
\end{center}}+e^4\,\Bigg[-
\parbox{18mm}{\begin{center}
\begin{fmfgraph}(15,5)
\setval
\fmfforce{0w,0.5h}{v1}
\fmfforce{1/3w,0.5h}{v2}
\fmfforce{1/2w,1h}{v2b}
\fmfforce{2/3w,0.5h}{v3}
\fmfforce{1/2w,0h}{v3b}
\fmfforce{1w,0.5h}{v4}
\fmf{boson}{v1,v2}
\fmf{fermion,right=0.45}{v2,v3b,v3,v2b,v2}
\fmf{boson}{v3,v4}
\fmf{boson}{v2b,v3b}
\fmfdot{v2,v2b,v3,v3b}
\end{fmfgraph}
\end{center}}-
\parbox{18mm}{\begin{center}
\begin{fmfgraph}(15,5)
\setval
\fmfforce{0w,0.5h}{v1}
\fmfforce{1/3w,0.5h}{v2}
\fmfforce{0.4008w,0.933h}{v2b}
\fmfforce{0.5992w,0.933h}{v2c}
\fmfforce{2/3w,0.5h}{v3}
\fmfforce{1w,0.5h}{v4}
\fmf{boson}{v1,v2}
\fmf{fermion,left=0.33}{v2,v2b,v2c,v3}
\fmf{fermion,left=1}{v3,v2}
\fmf{boson,right=0.8}{v2b,v2c}
\fmf{boson}{v3,v4}
\fmfdot{v2,v2b,v2c,v3}
\end{fmfgraph}
\end{center}}-
\parbox{18mm}{\begin{center}
\begin{fmfgraph}(15,5)
\setval
\fmfforce{0w,0.5h}{v1}
\fmfforce{1/3w,0.5h}{v2}
\fmfforce{0.4008w,0.933h}{v2b}
\fmfforce{0.5992w,0.933h}{v2c}
\fmfforce{2/3w,0.5h}{v3}
\fmfforce{1w,0.5h}{v4}
\fmf{boson}{v1,v2}
\fmf{fermion,right=0.33}{v3,v2c,v2b,v2}
\fmf{fermion,right=1}{v2,v3}
\fmf{boson,right=0.8}{v2b,v2c}
\fmf{boson}{v3,v4}
\fmfdot{v2,v2b,v2c,v3}
\end{fmfgraph}
\end{center}}\nonumber\\
&&+
\parbox{18mm}{\begin{center}
\begin{fmfgraph}(15,13)
\setval
\fmfforce{0w,0.192h}{v1}
\fmfforce{1/3w,0.192h}{v2}
\fmfforce{0.5w,0.384h}{v4}
\fmfforce{0.5w,0.615h}{v3}
\fmfforce{2/3w,0.192h}{v5}
\fmfforce{1w,0.192h}{v6}
\fmfi{fermion}{fullcircle rotated 270 scaled 1/3w shifted (0.5w,0.808h)}
\fmf{boson}{v1,v2}
\fmf{fermion,left=0.5}{v2,v4,v5}
\fmf{fermion,left=1}{v5,v2}
\fmf{boson}{v3,v4}
\fmf{boson}{v5,v6}
\fmfdot{v2,v3,v4,v5}
\end{fmfgraph}
\end{center}}+
\parbox{18mm}{\begin{center}
\begin{fmfgraph}(15,13)
\setval
\fmfforce{0w,0.192h}{v1}
\fmfforce{1/3w,0.192h}{v2}
\fmfforce{0.5w,0.384h}{v4}
\fmfforce{0.5w,0.615h}{v3}
\fmfforce{2/3w,0.192h}{v5}
\fmfforce{1w,0.192h}{v6}
\fmfi{fermion}{fullcircle rotated 270 scaled 1/3w shifted (0.5w,0.808h)}
\fmf{boson}{v1,v2}
\fmf{fermion,right=0.5}{v5,v4,v2}
\fmf{fermion,right=1}{v2,v5}
\fmf{boson}{v3,v4}
\fmf{boson}{v5,v6}
\fmfdot{v2,v3,v4,v5}
\end{fmfgraph}
\end{center}}+
\parbox{28mm}{\begin{center}
\begin{fmfgraph}(25,5)
\setval
\fmfforce{0w,0.5h}{v1}
\fmfforce{1/5w,0.5h}{v2}
\fmfforce{2/5w,0.5h}{v3}
\fmfforce{3/5w,0.5h}{v4}
\fmfforce{4/5w,0.5h}{v5}
\fmfforce{1w,0.5h}{v6}
\fmf{boson}{v1,v2}
\fmf{fermion,right}{v2,v3,v2}
\fmf{boson}{v3,v4}
\fmf{fermion,right}{v4,v5,v4}
\fmf{boson}{v5,v6}
\fmfdot{v2,v3,v4,v5}
\end{fmfgraph}
\end{center}}
\Bigg]+{\cal O}(e^6).
\end{eqnarray}
For brevity, we have omitted the labels $1$ and $2$
at the ends of the higher-order diagrams.
The full and the connected propagators 
${}^\gamma G^{2}_{12}$ and ${}^\gamma G^{2,c}_{12}$ satisfy the 
cumulant relation
\begin{equation}
  \label{np02b}
  {}^\gamma G^{2,c}_{12}={}^\gamma G^{2}_{12}-\mean{\hat{A}_1}\mean{\hat{A}_2}.
\end{equation}
Note that although the expectation value of the electromagnetic field
$\mean{\hat{A}_\mu ( x )}$ is zero in
quantum electrodynamics, it does not vanish in our generalized theory with
arbitrary 
propagators $S$ and $D$ [see Eq.~(\ref{vac14})].
 
The derivative of vacuum diagrams with respect to the electron kernel 
$S^{-1}$,
\begin{equation}
  \label{np03}
  {}^eG^{2}_{12}=\frac{\delta}{\delta S^{-1}_{21}}W[S^{-1},D^{-1},V],
\end{equation}
leads to the electronic two-point function, whose
diagrams are
\begin{eqnarray}
  \label{np04}
  {}^eG^{2}_{12}&\equiv& 
\parbox{22mm}{\begin{center}
\begin{fmfgraph*}(9,3)
\setval
\fmfleft{v1}
\fmfright{v2}
\fmf{fermion}{v2,v1}
\fmflabel{$\scs{1}$}{v1}
\fmflabel{$\scs{2}$}{v2}
\end{fmfgraph*}
\end{center}}+e^2\,\Bigg[
\parbox{18mm}{\begin{center}
\begin{fmfgraph}(15,5)
\setval
\fmfforce{0w,0.5h}{v1}
\fmfforce{1/3w,0.5h}{v2}
\fmfforce{2/3w,0.5h}{v3}
\fmfforce{1w,0.5h}{v4}
\fmf{fermion}{v4,v3,v2,v1}
\fmf{boson,left}{v2,v3}
\fmfdot{v2,v3}
\end{fmfgraph}
\end{center}}-
\parbox{13mm}{\begin{center}
\begin{fmfgraph}(10,8)
\setval
\fmfforce{0w,0h}{v1}
\fmfforce{1/2w,0h}{v2}
\fmfforce{1/2w,0.375h}{v2b}
\fmfforce{1w,0h}{v3}
\fmf{fermion}{v3,v2,v1}
\fmf{boson}{v2,v2b}
\fmfi{fermion}{fullcircle rotated 270 scaled 1/2w shifted (0.5w,0.6875h)}
\fmfdot{v2,v2b}
\end{fmfgraph}
\end{center}}
\Bigg]+e^4\Bigg[
\parbox{28mm}{\begin{center}
\begin{fmfgraph}(25,5)
\setval
\fmfforce{0w,0.5h}{v1}
\fmfforce{1/5w,0.5h}{v2}
\fmfforce{2/5w,0.5h}{v3}
\fmfforce{3/5w,0.5h}{v4}
\fmfforce{4/5w,0.5h}{v5}
\fmfforce{1w,0.5h}{v6}
\fmf{fermion}{v6,v5,v4,v3,v2,v1}
\fmf{boson,left=0.6}{v2,v4}
\fmf{boson,right=0.6}{v3,v5}
\fmfdot{v2,v3,v4,v5}
\end{fmfgraph}
\end{center}}+
\parbox{28mm}{\begin{center}
\begin{fmfgraph}(25,5)
\setval
\fmfforce{0w,0.5h}{v1}
\fmfforce{1/5w,0.5h}{v2}
\fmfforce{2/5w,0.5h}{v3}
\fmfforce{3/5w,0.5h}{v4}
\fmfforce{4/5w,0.5h}{v5}
\fmfforce{1w,0.5h}{v6}
\fmf{fermion}{v6,v5,v4,v3,v2,v1}
\fmf{boson,left}{v2,v3}
\fmf{boson,left}{v4,v5}
\fmfdot{v2,v3,v4,v5}
\end{fmfgraph}
\end{center}}\nonumber\\
&&+
\parbox{28mm}{\begin{center}
\begin{fmfgraph}(25,5)
\setval
\fmfforce{0w,0.5h}{v1}
\fmfforce{1/5w,0.5h}{v2}
\fmfforce{2/5w,0.5h}{v3}
\fmfforce{3/5w,0.5h}{v4}
\fmfforce{4/5w,0.5h}{v5}
\fmfforce{1w,0.5h}{v6}
\fmf{fermion}{v6,v5,v4,v3,v2,v1}
\fmf{boson,left=0.6}{v2,v5}
\fmf{boson,left=0.6}{v3,v4}
\fmfdot{v2,v3,v4,v5}
\end{fmfgraph}
\end{center}}-
\parbox{23mm}{\begin{center}
\begin{fmfgraph}(20,11)
\setval
\fmfforce{0w,3/11h}{v1}
\fmfforce{1/4w,3/11h}{v2}
\fmfforce{1/2w,3/11h}{v3}
\fmfforce{1/2w,6/11h}{v3b}
\fmfforce{3/4w,3/11h}{v4}
\fmfforce{1w,3/11h}{v5}
\fmf{fermion}{v5,v4,v3,v2,v1}
\fmf{boson,right=0.6}{v2,v4}
\fmf{boson}{v3,v3b}
\fmfi{fermion}{fullcircle rotated 270 scaled 1/4w shifted (0.5w,17/22h)}
\fmfdot{v2,v3,v3b,v4}
\end{fmfgraph}
\end{center}}-
\parbox{23mm}{\begin{center}
\begin{fmfgraph}(20,8)
\setval
\fmfforce{0w,0h}{v1}
\fmfforce{1/4w,0h}{v2}
\fmfforce{1/2w,0h}{v3}
\fmfforce{3/4w,0h}{v4}
\fmfforce{3/4w,3/8h}{v4b}
\fmfforce{1w,0h}{v5}
\fmf{fermion}{v5,v4,v3,v2,v1}
\fmf{boson,left}{v2,v3}
\fmf{boson}{v4,v4b}
\fmfi{fermion}{fullcircle rotated 270 scaled 1/4w shifted (0.75w,11/16h)}
\fmfdot{v2,v3,v4,v4b}
\end{fmfgraph}
\end{center}}-
\parbox{23mm}{\begin{center}
\begin{fmfgraph}(20,8)
\setval
\fmfforce{0w,0h}{v1}
\fmfforce{1/4w,0h}{v2}
\fmfforce{1/2w,0h}{v3}
\fmfforce{3/4w,0h}{v4}
\fmfforce{1/4w,3/8h}{v2b}
\fmfforce{1w,0h}{v5}
\fmf{fermion}{v5,v4,v3,v2,v1}
\fmf{boson,left}{v3,v4}
\fmf{boson}{v2,v2b}
\fmfi{fermion}{fullcircle rotated 270 scaled 1/4w shifted (0.25w,11/16h)}
\fmfdot{v2,v3,v4,v2b}
\end{fmfgraph}
\end{center}}\nonumber\\
&&-
\parbox{20mm}{\begin{center}
\begin{fmfgraph}(17,8)
\setval
\fmfforce{0w,0h}{v1}
\fmfforce{5/17w,0h}{v2}
\fmfforce{5/17w,3/8h}{v2b}
\fmfforce{12/17w,0h}{v3}
\fmfforce{12/17w,3/8h}{v3b}
\fmfforce{1w,0h}{v4}
\fmf{fermion}{v4,v3,v2,v1}
\fmf{boson}{v2,v2b}
\fmf{boson}{v3,v3b}
\fmfi{fermion}{fullcircle rotated 270 scaled 5/17w shifted (5/17w,11/16h)}
\fmfi{fermion}{fullcircle rotated 270 scaled 5/17w shifted (12/17w,11/16h)}
\fmfdot{v2,v3,v3b,v2b}
\end{fmfgraph}
\end{center}}-
\parbox{18mm}{\begin{center}
\begin{fmfgraph}(15,8)
\setval
\fmfforce{0w,0h}{v1}
\fmfforce{1/3w,0h}{v2}
\fmfforce{1/3w,11/16h}{v2b}
\fmfforce{2/3w,0h}{v3}
\fmfforce{2/3w,11/16h}{v3b}
\fmfforce{1w,0h}{v4}
\fmf{fermion}{v4,v3,v2,v1}
\fmf{boson}{v2,v2b}
\fmf{boson}{v3,v3b}
\fmf{fermion,right}{v2b,v3b,v2b}
\fmfdot{v2,v3,v3b,v2b}
\end{fmfgraph}
\end{center}}+
\parbox{13mm}{\begin{center}
\begin{fmfgraph}(10,16)
\setval
\fmfforce{0w,0h}{v1}
\fmfforce{1/2w,0h}{v2}
\fmfforce{1/2w,3/16h}{v2b}
\fmfforce{1/2w,1/2h}{v2c}
\fmfforce{1/2w,11/16h}{v2d}
\fmfforce{1w,0h}{v3}
\fmf{fermion}{v3,v2,v1}
\fmf{boson}{v2,v2b}
\fmf{boson}{v2c,v2d}
\fmf{fermion,right}{v2b,v2c,v2b}
\fmfi{fermion}{fullcircle rotated 270 scaled 1/2w shifted (1/2w,27/32h)}
\fmfdot{v2,v2b,v2c,v2d}
\end{fmfgraph}
\end{center}}-
\parbox{13mm}{\begin{center}
\begin{fmfgraph}(10,8)
\setval
\fmfforce{0w,0h}{v1}
\fmfforce{1/2w,0h}{v2}
\fmfforce{1/2w,3/8h}{v2b}
\fmfforce{0.2835w,0.84375h}{v2c}
\fmfforce{0.7165w,0.84375h}{v2d}
\fmfforce{1w,0h}{v3}
\fmf{fermion}{v3,v2,v1}
\fmf{boson}{v2,v2b}
\fmf{boson,right=0.3}{v2c,v2d}
\fmf{fermion,right=0.6}{v2b,v2d,v2c,v2b}
\fmfdot{v2,v2b,v2c,v2d}
\end{fmfgraph}
\end{center}}
\Bigg]+{\cal O}(e^6).
\end{eqnarray}
%
\subsection{Scattering Processes}
The generation of diagrams for scattering processes between electrons 
and photons (\ref{np00b}) and higher even-point functions
is now straightforward.
\subsubsection{Photon-Photon-Scattering}
The four-point function of photons is obtained by cutting two photon
lines in the vacuum diagrams or one photon line in the 
photonic two-point function: 
\begin{equation}
  \label{np09}
  {}^{\gamma\gamma}G^{4}_{1234}=4 \left\{ \frac{\delta^2 W}{\delta D^{-1}_{12} 
\delta D^{-1}_{34}} + 
\frac{\delta W}{\delta D^{-1}_{12}}
\frac{\delta W}{\delta D^{-1}_{34}} \right\}
=-2\frac{\delta \;{}^\gamma G^{2}_{12}}{\delta D^{-1}_{34}} +
{}^\gamma G^{2}_{12} {}^\gamma G^{2}_{34} .
\end{equation}
After applying one of the two possible operations in (\ref{np09}), 
the resulting connected diagrams to order $e^4$ are
\begin{equation}
  \label{np10}
  {}^{\gamma\gamma}G^{4,c}_{1234}\equiv -e^4\,\Bigg[
\parbox{28mm}{\begin{center}
\begin{fmfgraph*}(13,6)
\setval
\fmfforce{0w,1h}{i1}
\fmfforce{0w,0h}{i2}
\fmfforce{1w,1h}{o1}
\fmfforce{1w,0h}{o2}
\fmfforce{0.364w,0.2053h}{v1}
\fmfforce{0.364w,0.7957h}{v2}
\fmfforce{0.636w,0.7957h}{v3}
\fmfforce{0.636w,0.2053h}{v4}
\fmf{boson}{i1,v2}
\fmf{boson}{i2,v1}
\fmf{boson}{v3,o1}
\fmf{boson}{v4,o2}
\fmf{fermion,right=0.46}{v1,v4,v3,v2,v1}
\fmfdot{v1,v2,v3,v4}
\fmflabel{$\scs{2}$}{i1}
\fmflabel{$\scs{1}$}{i2}
\fmflabel{$\scs{3}$}{o1}
\fmflabel{$\scs{4}$}{o2}
\end{fmfgraph*}
\end{center}}+\,5\;\mbox{perm.}\Bigg]+{\cal O}(e^6),
\end{equation}
each permutation of two external spacetime coordinates leading to a 
different diagram. 
\subsubsection{M{\o}ller and Bhabba Scattering}
The scattering of two electrons (M{\o}ller scattering) is described 
by the electronic four-point 
function
\begin{equation}
  \label{np11}
  {}^{ee}G^{4}_{1234}=\frac{\delta^2 W}{\delta S^{-1}_{41}\delta 
S^{-1}_{32}} +\frac{\delta W}{\delta S^{-1}_{41} }
\frac{\delta W}{\delta S^{-1}_{32} }
=\frac{\delta \;{}^eG^{2}_{23}}{\delta S^{-1}_{41}}+
{}^eG^{2}_{14} {}^eG^{2}_{23}.
\end{equation}
To order $e^4$, the connected diagrams contributing to the 
fermionic four-point function are
\begin{eqnarray}
  \label{np12}
  {}^{ee}G^{4,c}_{1234}&\equiv& e^2\,\Bigg[
\parbox{28mm}{\begin{center}
\begin{fmfgraph*}(13,6)
\setval
\fmfleft{i2,i1}
\fmfright{o2,o1}
\fmf{fermion}{o2,v2,i2}
\fmf{fermion}{o1,v1,i1}
\fmf{boson}{v1,v2}
\fmfdot{v1,v2}
\fmflabel{$\scs{2}$}{i1}
\fmflabel{$\scs{1}$}{i2}
\fmflabel{$\scs{3}$}{o1}
\fmflabel{$\scs{4}$}{o2}
\end{fmfgraph*}
\end{center}}-\,({\scs 3}\leftrightarrow {\scs 4})\,\Bigg]+e^4\,\Bigg[
\parbox{18mm}{\begin{center}
\begin{fmfgraph}(13,8)
\setval
\fmfleft{i2,i1}
\fmfright{o2,o1}
\fmf{fermion,tension=2}{o1,v3,v1,v4,i1}
\fmf{fermion}{o2,v2,i2}
\fmf{boson}{v1,v2}
\fmffreeze
\fmf{boson,right=0.3}{v3,v4}
\fmfdot{v1,v2,v3,v4}
\end{fmfgraph}
\end{center}}+
\parbox{18mm}{\begin{center}
\begin{fmfgraph}(13,8)
\setval
\fmfleft{i2,i1}
\fmfright{o2,o1}
\fmf{fermion,tension=2}{o2,v3,v2,v4,i2}
\fmf{fermion}{o1,v1,i1}
\fmf{boson}{v1,v2}
\fmffreeze
\fmf{boson,left=0.3}{v3,v4}
\fmfdot{v1,v2,v3,v4}
\end{fmfgraph}
\end{center}}+
\parbox{18mm}{\begin{center}
\begin{fmfgraph}(13,8)
\setval
\fmfleft{i2,i1}
\fmfright{o2,o1}
\fmf{fermion,tension=2}{o1,v1,v2,i1}
\fmf{fermion,tension=2}{o2,v3,v4,i2}
\fmf{boson}{v1,v3}
\fmf{boson}{v2,v4}
\fmfdot{v1,v2,v3,v4}
\end{fmfgraph}
\end{center}}\nonumber\\
&&+
\parbox{18mm}{\begin{center}
\begin{fmfgraph}(13,8)
\setval
\fmfleft{i2,i1}
\fmfright{o2,o1}
\fmf{fermion,tension=2}{o1,v1,v2,i1}
\fmf{fermion,tension=2}{o2,v3,v4,i2}
\fmf{boson}{v1,v4}
\fmf{boson}{v2,v3}
\fmfdot{v1,v2,v3,v4}
\end{fmfgraph}
\end{center}}+
\parbox{18mm}{\begin{center}
\begin{fmfgraph}(13,8)
\setval
\fmfleft{i2,i1}
\fmfright{o2,o1}
\fmf{fermion}{o1,v1,i1}
\fmf{fermion,tension=1.5}{o2,v2,v3,v4,i2}
\fmf{boson,right=1,tension=0.4}{v3,v4}
\fmf{boson}{v1,v2}
\fmfdot{v1,v2,v3,v4}
\end{fmfgraph}
\end{center}}+
\parbox{18mm}{\begin{center}
\begin{fmfgraph}(13,8)
\setval
\fmfleft{i2,i1}
\fmfright{o2,o1}
\fmf{fermion}{o1,v1,i1}
\fmf{fermion,tension=1.5}{o2,v3,v4,v2,i2}
\fmf{boson,right=1,tension=0.4}{v3,v4}
\fmf{boson}{v1,v2}
\fmfdot{v1,v2,v3,v4}
\end{fmfgraph}
\end{center}}+
\parbox{18mm}{\begin{center}
\begin{fmfgraph}(13,8)
\setval
\fmfleft{i2,i1}
\fmfright{o2,o1}
\fmf{fermion,tension=1.5}{o1,v2,v3,v4,i1}
\fmf{fermion}{o2,v1,i2}
\fmf{boson,left=1,tension=0.4}{v3,v4}
\fmf{boson}{v1,v2}
\fmfdot{v1,v2,v3,v4}
\end{fmfgraph}
\end{center}}+
\parbox{18mm}{\begin{center}
\begin{fmfgraph}(13,8)
\setval
\fmfleft{i2,i1}
\fmfright{o2,o1}
\fmf{fermion,tension=1.5}{o1,v2,v3,v4,i1}
\fmf{fermion}{o2,v1,i2}
\fmf{boson,left=1,tension=0.4}{v2,v3}
\fmf{boson}{v1,v4}
\fmfdot{v1,v2,v3,v4}
\end{fmfgraph}
\end{center}}-
\parbox{18mm}{\begin{center}
\begin{fmfgraph}(13,8)
\setval
\fmfleft{i2,i1}
\fmfright{o2,o1}
\fmfforce{2/3w,0.2h}{v2}
\fmfforce{1/3w,0.1h}{v3}
\fmfforce{0.3w,0.35h}{v4}
\fmf{fermion,tension=2}{o1,v1,i1}
\fmf{fermion,tension=1.5}{o2,v2,v3,i2}
\fmf{boson}{v3,v4}
\fmfi{fermion}{fullcircle rotated -90 scaled 5/26w shifted (0.285w,0.5h)}
\fmf{boson}{v1,v2}
\fmfdot{v1,v2,v3,v4}
\end{fmfgraph}
\end{center}}\nonumber\\
&&-
\parbox{18mm}{\begin{center}
\begin{fmfgraph}(13,8)
\setval
\fmfleft{i2,i1}
\fmfright{o2,o1}
\fmfforce{2/3w,0.1h}{v2}
\fmfforce{1/3w,0.2h}{v3}
\fmfforce{0.7w,0.35h}{v4}
\fmf{fermion,tension=2}{o1,v1,i1}
\fmf{fermion,tension=1.5}{o2,v2,v3,i2}
\fmf{boson}{v2,v4}
\fmfi{fermion}{fullcircle rotated -90 scaled 5/26w shifted (0.715w,0.5h)}
\fmf{boson}{v1,v3}
\fmfdot{v1,v2,v3,v4}
\end{fmfgraph}
\end{center}}-
\parbox{18mm}{\begin{center}
\begin{fmfgraph}(13,8)
\setval
\fmfleft{i2,i1}
\fmfright{o2,o1}
\fmfforce{1/3w,0.9h}{v2}
\fmfforce{2/3w,0.8h}{v3}
\fmfforce{0.3w,0.65h}{v4}
\fmf{fermion,tension=1.5}{o1,v3,v2,i1}
\fmf{fermion,tension=2}{o2,v1,i2}
\fmf{boson}{v2,v4}
\fmfi{fermion}{fullcircle rotated 90 scaled 5/26w shifted (0.285w,0.5h)}
\fmf{boson}{v1,v3}
\fmfdot{v1,v2,v3,v4}
\end{fmfgraph}
\end{center}}-
\parbox{18mm}{\begin{center}
\begin{fmfgraph}(13,8)
\setval
\fmfleft{i2,i1}
\fmfright{o2,o1}
\fmfforce{2/3w,0.9h}{v2}
\fmfforce{1/3w,0.8h}{v3}
\fmfforce{0.7w,0.65h}{v4}
\fmf{fermion,tension=1.5}{o1,v2,v3,i1}
\fmf{fermion,tension=2}{o2,v1,i2}
\fmf{boson}{v2,v4}
\fmfi{fermion}{fullcircle rotated 90 scaled 5/26w shifted (0.715w,0.5h)}
\fmf{boson}{v1,v3}
\fmfdot{v1,v2,v3,v4}
\end{fmfgraph}
\end{center}}-
\parbox{18mm}{\begin{center}
\begin{fmfgraph}(13,8)
\setval
\fmfleft{i2,i1}
\fmfright{o2,o1}
\fmf{fermion}{o1,v1,i1}
\fmf{fermion}{o2,v4,i2}
\fmf{boson}{v1,v2}
\fmf{fermion,right=1,tension=0.4}{v2,v3,v2}
\fmf{boson}{v3,v4}
\fmfdot{v1,v2,v3,v4}
\end{fmfgraph}
\end{center}}
-\,({\scs 3}\leftrightarrow {\scs 4})\,\Bigg]+{\cal O}(e^6),
\end{eqnarray}
where the spacetime indices in all diagrams are arranged as in the 
first. Each diagram on the right-hand side has a partner 
with opposite sign, where the spacetime indices either of the incoming 
or of the outgoing electrons are interchanged. 
The tadpole diagrams vanish for physical
propagators $S=S_{\rm F}$, $D=D_{\rm F}$, and the corresponding corrections attached to
external legs do not contribute
when calculating the $S$-matrix elements. In our general 
vacuum functional, however, we 
must not discard them, since they contribute to higher functional
derivatives, which would be needed for the calculation of, e.g., the 
six-point function.

By interchanging spacetime arguments in the kernels of Eq.~(\ref{np12})
apparently, the Feynman diagrams (\ref{np12}) describe also
scattering of electron and positron (Bhabba scattering) 
and scattering of two positrons.
\subsubsection{Compton Scattering}
The amplitude of Compton scattering is given by the 
mixed four-point function ${}^{e\gamma}G^{4}_{1234}$.
To obtain the relevant 
Feynman diagrams, we have to perform one of the possible operations
\begin{eqnarray}
  \label{np13}
  {}^{e\gamma}G^{4}_{1234}=
-2 \left\{ \frac{\delta^2 W}{\delta D^{-1}_{23}\delta S^{-1}_{41}}
+\frac{\delta W}{\delta D^{-1}_{23}}
\frac{\delta W}{\delta S^{-1}_{41}} \right\}
=
\frac{\delta \;{}^\gamma G^{2}_{23}}{\delta S^{-1}_{41}}
+ {}^eG^{2}_{14} {}^\gamma G^{2}_{23}
=
-2\frac{\delta \;{}^eG^{2}_{14}}{\delta D^{-1}_{23}}
+ {}^eG^{2}_{14} {}^\gamma G^{2}_{23} .
\end{eqnarray}
The resulting connected Feynman diagrams to order $e^4$ are
\begin{eqnarray}
  \label{np14}
  {}^{e\gamma}G^{4,c}_{1234}&\equiv& e^2\,\Bigg[
\parbox{28mm}{\begin{center}
\begin{fmfgraph*}(13,6)
\setval
\fmfleft{i2,i1}
\fmfright{o2,o1}
\fmf{fermion}{o2,v2,v1,i2}
\fmf{boson}{o1,v2}
\fmf{boson}{i1,v1}
\fmfdot{v1,v2}
\fmflabel{$\scs{2}$}{i1}
\fmflabel{$\scs{1}$}{i2}
\fmflabel{$\scs{3}$}{o1}
\fmflabel{$\scs{4}$}{o2}
\end{fmfgraph*}
\end{center}}+\,({\scs 2}\leftrightarrow {\scs 3})\,\Bigg]+e^4\,\Bigg[
\parbox{18mm}{\begin{center}
\begin{fmfgraph}(13,8)
\setval
\fmfleft{i2,i1}
\fmfright{o2,o1}
\fmf{fermion,tension=2}{o2,v1,v2,v3,v4}
\fmf{fermion,tension=0.9}{v4,i2}
\fmf{boson}{o1,v2}
\fmf{boson}{i1,v4}
\fmffreeze
\fmf{boson,left=0.5}{v1,v3}
\fmfdot{v1,v2,v3,v4}
\end{fmfgraph}
\end{center}}+
\parbox{18mm}{\begin{center}
\begin{fmfgraph}(13,8)
\setval
\fmfleft{i2,i1}
\fmfright{o2,o1}
\fmf{fermion,tension=0.9}{o2,v1}
\fmf{fermion,tension=2}{v1,v2,v3,v4,i2}
\fmf{boson}{o1,v1}
\fmf{boson}{i1,v3}
\fmffreeze
\fmf{boson,left=0.5}{v2,v4}
\fmfdot{v1,v2,v3,v4}
\end{fmfgraph}
\end{center}}+
\parbox{18mm}{\begin{center}
\begin{fmfgraph}(13,8)
\setval
\fmfleft{i2,i1}
\fmfright{o2,o1}
\fmf{fermion}{o2,v1}
\fmf{fermion,tension=2}{v1,v2,v3,v4}
\fmf{fermion}{v4,i2}
\fmf{boson}{o1,v1}
\fmf{boson}{i1,v4}
\fmffreeze
\fmf{boson,right=1.5}{v2,v3}
\fmfdot{v1,v2,v3,v4}
\end{fmfgraph}
\end{center}}\nonumber\\
&&+
\parbox{18mm}{\begin{center}
\begin{fmfgraph}(13,8)
\setval
\fmfleft{i2,i1}
\fmfright{o2,o1}
\fmf{fermion,tension=2}{o2,v1,v2,v3}
\fmf{fermion,tension=2}{v3,v4}
\fmf{fermion,tension=0.66}{v4,i2}
\fmf{boson}{o1,v3}
\fmf{boson}{i1,v4}
\fmffreeze
\fmf{boson,left=1.5}{v1,v2}
\fmfdot{v1,v2,v3,v4}
\end{fmfgraph}
\end{center}}+
\parbox{18mm}{\begin{center}
\begin{fmfgraph}(13,8)
\setval
\fmfleft{i2,i1}
\fmfright{o2,o1}
\fmf{fermion,tension=2}{v2,v3,v4,i2}
\fmf{fermion,tension=2}{v1,v2}
\fmf{fermion,tension=0.66}{o2,v1}
\fmf{boson}{o1,v1}
\fmf{boson}{i1,v2}
\fmffreeze
\fmf{boson,left=1.5}{v3,v4}
\fmfdot{v1,v2,v3,v4}
\end{fmfgraph}
\end{center}}+
\parbox{18mm}{\begin{center}
\begin{fmfgraph}(13,8)
\setval
\fmfleft{i2,i1}
\fmfright{o2,o1}
\fmf{fermion,tension=2}{v4,i2}
\fmf{fermion}{v3,v4}
\fmf{fermion,tension=2}{v2,v3}
\fmf{fermion}{v1,v2}
\fmf{fermion,tension=2}{o2,v1}
\fmf{boson,tension=2}{o1,v2}
\fmf{boson,tension=2}{i1,v3}
\fmf{boson,tension=2}{v1,v4}
\fmfdot{v1,v2,v3,v4}
\end{fmfgraph}
\end{center}}-
\parbox{18mm}{\begin{center}
\begin{fmfgraph}(13,8)
\setval
\fmfleft{i2,i1}
\fmfright{o2,o1}
\fmfforce{0.5w,0.5h}{v2}
\fmfforce{0.5w,0.7h}{v4}
\fmf{fermion}{o2,v1}
\fmf{fermion,tension=2}{v1,v2,v3}
\fmf{fermion}{v3,i2}
\fmf{boson}{o1,v1}
\fmf{boson}{i1,v3}
\fmf{boson}{v2,v4}
\fmfi{fermion}{fullcircle rotated -90 scaled 5/26w shifted (0.5w,0.85h)}
\fmfdot{v1,v2,v3,v4}
\end{fmfgraph}
\end{center}}-
\parbox{18mm}{\begin{center}
\begin{fmfgraph}(13,8)
\setval
\fmfleft{i2,i1}
\fmfright{o2,o1}
\fmfforce{0.78w,0.22h}{v1}
\fmfforce{0.5w,0.5h}{v2}
\fmfforce{0.25w,0.5h}{v3}
\fmfforce{0.85w,0.4h}{v4}
\fmf{fermion}{o2,v1,v2}
\fmf{fermion}{v2,v3}
\fmf{fermion}{v3,i2}
\fmf{boson}{o1,v2}
\fmf{boson}{i1,v3}
\fmf{boson}{v1,v4}
\fmfi{fermion}{fullcircle rotated -135 scaled 5/26w shifted (0.9w,0.55h)}
\fmfdot{v1,v2,v3,v4}
\end{fmfgraph}
\end{center}}-
\parbox{18mm}{\begin{center}
\begin{fmfgraph}(13,8)
\setval
\fmfleft{i2,i1}
\fmfright{o2,o1}
\fmfforce{0.22w,0.22h}{v1}
\fmfforce{0.5w,0.5h}{v2}
\fmfforce{0.75w,0.5h}{v3}
\fmfforce{0.15w,0.4h}{v4}
\fmf{fermion}{o2,v3}
\fmf{fermion}{v3,v2}
\fmf{fermion}{v2,v1,i2}
\fmf{boson}{o1,v3}
\fmf{boson}{i1,v2}
\fmf{boson}{v1,v4}
\fmfi{fermion}{fullcircle rotated -45 scaled 5/26w shifted (0.1w,0.55h)}
\fmfdot{v1,v2,v3,v4}
\end{fmfgraph}
\end{center}}\nonumber\\
&&-
\parbox{18mm}{\begin{center}
\begin{fmfgraph}(13,8)
\setval
\fmfleft{i2,i1}
\fmfright{o2,o1}
\fmfforce{0.55w,0.5h}{v1}
\fmfforce{0.25w,0.5h}{v2}
\fmf{fermion}{o2,v1,v2,i2}
\fmf{boson}{o1,v3}
\fmf{fermion,right=1,tension=0.45}{v3,v4,v3}
\fmf{boson}{v4,v1}
\fmf{boson}{i1,v2}
\fmfdot{v1,v2,v3,v4}
\end{fmfgraph}
\end{center}}-
\parbox{18mm}{\begin{center}
\begin{fmfgraph}(13,8)
\setval
\fmfleft{i2,i1}
\fmfright{o2,o1}
\fmfforce{0.45w,0.5h}{v1}
\fmfforce{0.75w,0.5h}{v2}
\fmf{fermion}{o2,v2,v1,i2}
\fmf{boson}{i1,v3}
\fmf{fermion,right=1,tension=0.45}{v3,v4,v3}
\fmf{boson}{v4,v1}
\fmf{boson}{o1,v2}
\fmfdot{v1,v2,v3,v4}
\end{fmfgraph}
\end{center}}-
\parbox{18mm}{\begin{center}
\begin{fmfgraph}(13,8)
\setval
\fmfleft{i2,i1}
\fmfright{o2,o1}
\fmf{fermion}{o2,v1,i2}
\fmf{boson}{i1,v3}
\fmf{fermion,right=0.3}{v3,v2,v4}
\fmf{fermion,right=1}{v4,v3}
\fmf{boson}{o1,v4}
\fmf{boson}{v1,v2}
\fmfdot{v1,v2,v3,v4}
\end{fmfgraph}
\end{center}}
+ ({\scs 2}\leftrightarrow {\scs 3})\Bigg]+{\cal O}(e^6),
\end{eqnarray}
where the diagrams with interchanged photon coordinates 
$2\leftrightarrow 3$ possess the same sign as the original one. 
%
%
\subsection{Three-Point Vertex Function}
The three-point vertex function is obtained from the vacuum energy 
$W$ by performing the derivative with respect to the interaction function
$V_{123}$, which we have defined in Eq.~(\ref{vert00}):
\begin{equation}
  \label{np16}
  G^3_{123}=-\frac{1}{e}\frac{\delta W}{\delta V_{213}}.
\end{equation}
The easiest way to find the associated Feynman diagrams is to
apply the graphical operation (\ref{vert01}), which removes a vertex
from the vacuum diagrams in all possible ways 
and lets the remaining legs open.  
Dropping disconnected diagrams by considering the cumulant
\begin{equation}
  \label{np17}
  G^{3,c}_{123}=G^{3}_{123}-\mean{\hat{\psi}_1\,\hat{\bar{\psi}}_2}
\mean{\hat{A}_3}
\end{equation}
we obtain
\begin{equation}
  \label{np18}
  G^{3,c}_{123}=\;e
\parbox{17mm}{\centerline{
\begin{fmfgraph*}(10,8.66)
\setval
\fmfforce{1w,0h}{v1}
\fmfforce{0w,0h}{v2}
\fmfforce{0.5w,1h}{v3}
\fmfforce{0.5w,0.2886h}{vm}
\fmf{fermion}{v1,vm,v2}
\fmf{boson}{v3,vm}
\fmfv{decor.size=0,label={\footnotesize 2},l.dist=0.5mm}{v1}
\fmfv{decor.size=0,label={\footnotesize 1},l.dist=0.5mm}{v2}
\fmfv{decor.size=0,label={\footnotesize 3},l.dist=0.5mm}{v3}
\fmfdot{vm}
\end{fmfgraph*}
}}+\;e^3\,\Bigg[
\parbox{15mm}{\centerline{
\begin{fmfgraph}(10,8.66)
\setval
\fmfforce{1w,0h}{v1}
\fmfforce{0.75w,0.225h}{vm1}
\fmfforce{0w,0h}{v2}
\fmfforce{0.25w,0.225h}{vm2}
\fmfforce{0.5w,1h}{v3}
\fmfforce{0.5w,0.45h}{vm}
\fmf{fermion}{v1,vm1,vm,vm2,v2}
\fmf{boson,right=0.4}{vm2,vm1}
\fmf{boson}{v3,vm}
\fmfdot{vm,vm1,vm2}
\end{fmfgraph}}}+
\parbox{15mm}{\centerline{
\begin{fmfgraph}(10,8.66)
\setval
\fmfforce{1w,0h}{v1}
\fmfforce{0.833w,0.15h}{vm1}
\fmfforce{0.667w,0.3h}{vm2}
\fmfforce{0w,0h}{v2}
\fmfforce{0.5w,1h}{v3}
\fmfforce{0.5w,0.45h}{vm}
\fmf{fermion}{v1,vm1,vm2,vm,v2}
\fmf{boson,right=1.2}{vm1,vm2}
\fmf{boson}{v3,vm}
\fmfdot{vm,vm1,vm2}
\end{fmfgraph}}}+
\parbox{15mm}{\centerline{
\begin{fmfgraph}(10,8.66)
\setval
\fmfforce{1w,0h}{v1}
\fmfforce{0.333w,0.3h}{vm1}
\fmfforce{0.167w,0.15h}{vm2}
\fmfforce{0w,0h}{v2}
\fmfforce{0.5w,1h}{v3}
\fmfforce{0.5w,0.45h}{vm}
\fmf{fermion}{v1,vm,vm1,vm2,v2}
\fmf{boson,right=1.2}{vm1,vm2}
\fmf{boson}{v3,vm}
\fmfdot{vm,vm1,vm2}
\end{fmfgraph}}}-
\parbox{15mm}{\centerline{
\begin{fmfgraph}(10,8.66)
\setval
\fmfforce{1w,0h}{v1}
\fmfforce{0.75w,0.225h}{vm1}
\fmfforce{0.85w,0.37h}{vm2}
\fmfforce{0w,0h}{v2}
\fmfforce{0.5w,1h}{v3}
\fmfforce{0.5w,0.45h}{vm}
\fmf{fermion}{v1,vm1,vm,v2}
\fmfi{fermion}{fullcircle rotated -135 scaled 1/5w shifted (0.92w,0.46h)}
\fmf{boson}{vm2,vm1}
\fmf{boson}{v3,vm}
\fmfdot{vm,vm1,vm2}
\end{fmfgraph}}}-
\parbox{15mm}{\centerline{
\begin{fmfgraph}(10,8.66)
\setval
\fmfforce{1w,0h}{v1}
\fmfforce{0.25w,0.225h}{vm1}
\fmfforce{0.15w,0.37h}{vm2}
\fmfforce{0w,0h}{v2}
\fmfforce{0.5w,1h}{v3}
\fmfforce{0.5w,0.45h}{vm}
\fmf{fermion}{v1,vm,vm1,v2}
\fmfi{fermion}{fullcircle rotated -45 scaled 1/5w shifted 
(0.08w,0.46h)}
\fmf{boson}{vm2,vm1}
\fmf{boson}{v3,vm}
\fmfdot{vm,vm1,vm2}
\end{fmfgraph}}}-
\parbox{15mm}{\centerline{
\begin{fmfgraph}(10,8.66)
\setval
\fmfforce{1w,0h}{v1}
\fmfforce{0w,0h}{v2}
\fmfforce{0.5w,1h}{v3}
\fmfforce{0.5w,0.745h}{vm1}
\fmfforce{0.5w,0.455h}{vm2}
\fmfforce{0.5w,0.2h}{vm}
\fmf{fermion}{v1,vm,v2}
\fmf{boson}{vm,vm2}
\fmf{boson}{vm1,v3}
\fmf{fermion,right=1}{vm2,vm1,vm2}
\fmfdot{vm,vm1,vm2}
\end{fmfgraph}}}
\Bigg]+{\cal O}(e^5).
\end{equation}
\end{fmffile}
\begin{fmffile}{diag3bkpB}
\section{Scattering of Electrons and Photons in the Presence of an External
Electromagnetic Field}
\label{scatj}
To describe the scattering of electrons and photons
on external electromagnetic fields, the action 
${\cal A}[\bar{\psi},\psi,A]$ in Eq.~(\ref{gt18b}) must be extended 
by an additional
external current $J$, which is coupled linearly to the 
electromagnetic field $A$:
\begin{equation}
  \label{scj00}
  {\cal A}^J[\bar{\psi},\psi,A,J]={\cal A}[\bar{\psi},\psi,A]-
e\int_1 J_1 A_1.
\end{equation}
Then the partition function (\ref{gt18}) becomes a functional in the 
physical current $J$ and is given by
\begin{equation}
  \label{scj01}
  Z[J]=\oint\meas {\rm e}^{-{\cal A}^J[\bar{\psi},\psi,A,J]}
\end{equation}
with $Z=Z[0]$.
The external current is usually supplied by some atomic nucleus of charge
$N e$ with integer number $N$. For this reason, the factor $e$ is removed 
from the current in Eq.~(\ref{scj00}) to be
able to 
collect systematically all Feynman diagrams
of the same order in $e$. This organization may not always be the most useful
one. If we consider, for instance, an external heavy nucleus with a high
charge $N e$, we may have to include many more orders in the external charge
$N e$ than in the internal charge $e$. Such subleties will be ignored here, 
for simplicity.
\subsection{Recursion Relation for the Vacuum Energy with External Source}
Along similar lines as before, we derive
the recursion relation for the vacuum energy in the presence of an
external current, 
$W[J]={\rm ln}\,Z[J]$ which is 
now also a functional of $J$ (suppressing the other arguments $D^{-1}, S^{-1},
V$). 
After that, we derive a recursion relation
{\it only} producing those vacuum diagrams which contain a coupling to the source. 
It turns out that the resulting recursion relation for current diagrams
is extremly simple. Hence, this recursion relation 
is the ideal 
extension of the former Eq.~(\ref{vac26}) which generates only the 
source-free diagrams.
\subsubsection{Complete Recursion Relation for All Vacuum Diagrams}
The recursion relation for {\it all} vacuum diagrams with and without
external source is derived in a similar manner as that for all 
source-free vacuum diagrams (\ref{vac26}).
There will be, however, a few significant differences
in comparison with the procedure in Sec.~\ref{recrel}. Since the current
$J$ couples to the electromagnetic field $A$, vacuum 
diagrams with external current always contain
photon lines. For this reason, we start with the identity
\begin{equation}
  \label{rec00}
  \oint \meas\frac{\delta}{\delta A_1}
\left\{A_2 {\rm e}^{-{\cal A}^J[\bar{\psi},\psi,A,J]}\right\}=0
\end{equation}
instead of Eq.~(\ref{vac11}). Performing the functional derivative leads to
\begin{equation}
  \label{rec01}
  Z[J]\delta_{12}+2\int_3 D^{-1}_{13}\frac{\delta Z[J]}
{\delta D^{-1}_{23}}
-e\int_{34}V_{341}\frac{\delta}{\delta S^{-1}_{34}}[\meanJ{\hat{A}_2} 
Z[J]]
+e J_1\meanJ{\hat{A}_2}Z[J]=0
\end{equation}
in analogy to Eq.~(\ref{vac12}). The expectation value of the electromagnetic
field $A$ in
the presence of an external source $J$ is found by exploiting the identity 
\begin{equation}
  \label{rec02}
  \oint\meas \frac{\delta}{\delta A_1}\,{\rm e}^{-{\cal A}^J
[\bar{\psi},\psi,A,J]}=0
\end{equation}
to derive, as in Eqs.~(\ref{vac13})--(\ref{vac14}),
\begin{equation}
  \label{rec03}
  \meanJ{\hat{A}_1}=-e\int_{234}V_{234}D_{14}\frac{\delta W[J]}
{\delta S^{-1}_{23}}+e\int_2 D_{12} J_2,
\end{equation}
where we have set $W[J]={\rm ln}\,Z[J]$. Inserting the
expectation value (\ref{rec03}) into Eq.~(\ref{rec01}), the resulting
functional differential equation reads
\begin{eqnarray}
  \label{rec04}
  \delta_{12}+2\int_3 D^{-1}_{13}\frac{\delta W[J]}
{\delta D^{-1}_{23}}&=&-e^2\int_{3\cdots 7}V_{341}V_{567}D_{27}\left\{
\frac{\delta^2W[J]}{\delta S^{-1}_{34}\delta S^{-1}_{56}}+
\frac{\delta W[J]}{\delta S^{-1}_{34}}
\frac{\delta W[J]}{\delta S^{-1}_{56}}\right\}\nonumber\\
&&\,+2e^2\int_{345}V_{345}
D_{25}J_1\frac{\delta W[J]}{\delta S^{-1}_{34}}-e^2\int_3 
J_1D_{23}J_3.
\end{eqnarray}
Using relations (\ref{rr19}) and (\ref{rr31}), and taking the trace,
this becomes
\begin{eqnarray}
  \label{rec05}
  -\int_1\delta_{11}+2\int_{12}D_{12}\frac{\delta W[J]}
{\delta D_{12}}&=&2e^2\int_{1\cdots 8}V_{123}V_{456}D_{36}S_{71}S_{24}S_{58}
\frac{\delta W[J]}{\delta S_{78}}+e^2\int_{1\cdots 10}V_{123}V_{456}
D_{36}S_{71}S_{28}S_{94}S_{5\,10}\nonumber\\
&&\hspace*{-80pt}\times\left\{ \frac{\delta^2W[J]}
{\delta S_{78}\delta S_{9\,10}}+
\frac{\delta W[J]}{\delta S_{78}}
\frac{\delta W[J]}{\delta S_{9\,10}}\right\}
+2e^2\int_{1\cdots 6}
V_{123}D_{34}J_4S_{51}S_{26}\frac{\delta W[J]}
{\delta S_{56}}+e^2\int_{12}J_1D_{12}J_2,
\end{eqnarray}
which generalizes Eq.~(\ref{vac15b}). Expanding $W[J]$ as in 
Eq.~(\ref{gt19f}),
\begin{equation}
\label{rec05b}
W[J]=W^{(0)}+\sum\limits_{p=1}^\infty e^{2p} \,W^{(p)}[J],
\end{equation} 
and using the fact that the free vacuum energy 
$W^{(0)}[J]=W^{(0)}[0]=W^{(0)}$ 
is independent of the external current, the first term on the left-hand 
side in Eq.~(\ref{rec05}) is canceled by an identity following from 
Eq.~(\ref{rr09})
\begin{equation}
\label{rec06}
2\int_{12}D_{12}\frac{\delta W^{(0)}}{\delta D_{12}}=\int_1\delta_{11}.
\end{equation}
Introducing a Feynman diagram for the coupling to the current $J$
\begin{equation}
  \label{rec06b}
\parbox{11mm}{\centerline{  
\begin{fmfgraph}(7,6)
\setval
\fmfleft{i1,i2}
\fmfright{o}
\fmfforce{0.5w,0.5h}{v}
\fmf{double,width=0.2mm}{i1,v,i2}
\fmf{boson}{v,o}
\fmfdot{v}
\end{fmfgraph}
}}{\scs 1}\quad\equiv\quad J_1,
\end{equation}
we obtain the graphical recursion relation
\begin{eqnarray}
  \label{rec07}
2\parbox{7mm}{\centerline{
\begin{fmfgraph*}(3,6)
\setval
\fmfforce{0.5w,1h}{i1}
\fmfforce{0.5w,0h}{i2}
\fmf{boson,right=0.6}{i1,i2}
\fmfv{decor.size=0, label=${\scs 1}$, l.dist=1mm, l.angle=0}{i1}
\fmfv{decor.size=0, label=${\scs 2}$, l.dist=1mm, l.angle=0}{i2}
\end{fmfgraph*}
}} \dbose{W^{(p+1)}[J]}{1}{2}\;&\equiv&\;
\parbox{14mm}{\begin{center}
\begin{fmfgraph*}(8,9)
\setval
\fmfstraight
\fmfforce{0w,0.835h}{i1}
\fmfforce{0w,0.165h}{i2}
\fmfforce{1w,1h}{o1}
\fmfforce{1w,0.66h}{o2}
\fmfforce{1w,0.33h}{o3}
\fmfforce{1w,0h}{o4}
\fmf{boson}{i1,i2}
\fmf{fermion,left=0.1}{i1,o1}
\fmf{fermion,left=0.1}{o2,i1}
\fmf{fermion,left=0.1}{i2,o3}
\fmf{fermion,left=0.1}{o4,i2}
\fmfdot{i1,i2}
\fmfv{decor.size=0, label=${\scs 1}$, l.dist=1mm, l.angle=0}{o1}
\fmfv{decor.size=0, label=${\scs 2}$, l.dist=1mm, l.angle=0}{o2}
\fmfv{decor.size=0, label=${\scs 3}$, l.dist=1mm, l.angle=0}{o3}
\fmfv{decor.size=0, label=${\scs 4}$, l.dist=1mm, l.angle=0}{o4}
\end{fmfgraph*}
\end{center}}
\quad \ddfermi{W^{(p)}[J]}\; +\;2\;\Bigg[
\parbox{13mm}{\begin{center}
\begin{fmfgraph*}(8,6)
\setval
\fmfstraight
\fmfforce{0.3w,1h}{i1}
\fmfforce{0.3w,0h}{i2}
\fmfforce{1w,1h}{o1}
\fmfforce{1w,0h}{o2}
\fmf{fermion}{i2,i1}
\fmf{boson,right=0.7}{i1,i2}
\fmf{fermion}{i1,o1}
\fmf{fermion}{o2,i2}
\fmfdot{i1,i2}
\fmfv{decor.size=0, label=${\scs 1}$, l.dist=1mm, l.angle=0}{o1}
\fmfv{decor.size=0, label=${\scs 2}$, l.dist=1mm, l.angle=0}{o2}
\end{fmfgraph*}
\end{center}}
\quad-\;
\parbox{17mm}{\begin{center}
\begin{fmfgraph*}(12,6)
\setval
\fmfstraight
\fmfforce{0.33w,0.5h}{v1}
\fmfforce{0.66w,0.5h}{v2}
\fmfforce{1w,1h}{o1}
\fmfforce{1w,0h}{o2}
\fmf{boson}{v1,v2}
\fmf{fermion}{v2,o1}
\fmf{fermion}{o2,v2}
\fmfi{fermion}{reverse fullcircle scaled 0.33w shifted (0.165w,0.5h)}
\fmfdot{v1,v2}
\fmfv{decor.size=0, label=${\scs 1}$, l.dist=1mm, l.angle=0}{o1}
\fmfv{decor.size=0, label=${\scs 2}$, l.dist=1mm, l.angle=0}{o2}
\end{fmfgraph*}
\end{center}}
\quad\Bigg] \dfermi{W^{(p)}[J]}{1}{2}\nonumber\\
&&\hspace*{-20pt}+\;\sum\limits_{q=1}^{p-1}\quad
\dfermi{W^{(p-q)}[J]}{1}{2}\quad
\parbox{17mm}{\begin{center}
\begin{fmfgraph*}(12,6)
\setval
\fmfstraight
\fmfleft{i2,i1}
\fmfright{o2,o1}
\fmf{fermion}{v1,i1}
\fmf{fermion}{i2,v1}
\fmf{boson}{v1,v2}
\fmf{fermion}{v2,o1}
\fmf{fermion}{o2,v2}
\fmfdot{v1,v2}
\fmfv{decor.size=0, label=${\scs 1}$, l.dist=1mm, l.angle=-180}{i1}
\fmfv{decor.size=0, label=${\scs 2}$, l.dist=1mm, l.angle=-180}{i2}
\fmfv{decor.size=0, label=${\scs 3}$, l.dist=1mm, l.angle=0}{o1}
\fmfv{decor.size=0, label=${\scs 4}$, l.dist=1mm, l.angle=0}{o2}
\end{fmfgraph*}
\end{center}}
\quad\dfermi{W^{(q)}[J]}{3}{4}+2
\parbox{17mm}{\begin{center}
\begin{fmfgraph*}(12,6)
\setval
\fmfstraight
\fmfleft{i2,i1}
\fmfright{o2,o1}
\fmf{double,width=0.2mm}{i2,v1,i1}
\fmf{boson}{v1,v2}
\fmf{fermion}{v2,o1}
\fmf{fermion}{o2,v2}
\fmfdot{v1,v2}
\fmfv{decor.size=0, label=${\scs 1}$, l.dist=1mm, l.angle=0}{o1}
\fmfv{decor.size=0, label=${\scs 2}$, l.dist=1mm, l.angle=0}{o2}
\end{fmfgraph*}
\end{center}}
\quad \dfermi{W^{(p)}[J]}{1}{2}
,\qquad p\ge 1,
\end{eqnarray}
and the first-order diagrams
\begin{equation}
  \label{rec08}
  W^{(1)}[J]=W^{(1)}[0]\;+\;\frac{1}{2}
\parbox{14mm}{\begin{center}
\begin{fmfgraph}(10,6)
\setval
\fmfforce{0w,0h}{i1}
\fmfforce{3/10w,0.5h}{i2}
\fmfforce{0w,1h}{i3}
\fmfforce{1w,0h}{o1}
\fmfforce{7/10w,0.5h}{o2}
\fmfforce{1w,1h}{o3}
\fmf{double,width=0.2mm}{i1,i2,i3}
\fmf{double,width=0.2mm}{o1,o2,o3}
\fmf{boson}{i2,o2}
\fmfdot{i2,o2}
\end{fmfgraph}
\end{center}}-
\parbox{16mm}{\begin{center}
\begin{fmfgraph}(12,6)
\setval
\fmfleft{i1,i2}
\fmfforce{1/4w,0.5h}{v1}
\fmfforce{7/12w,0.5h}{v2}
\fmf{double,width=0.2mm}{i1,v1,i2}
\fmf{boson}{v1,v2}
\fmfi{fermion}{fullcircle rotated 180 scaled 5/12w shifted (0.792w,0.5h)}
\fmfdot{v1,v2}
\end{fmfgraph}
\end{center}},
\end{equation}
where $W^{(1)}[0]=W^{(1)}$ contains the source-free first-order vacuum 
diagrams 
(\ref{vac06}). An important difference between the recursion relation 
(\ref{rec07}) and the previous 
(\ref{vac26}) is that the vacuum diagrams in a series of
the coupling constant $e$ contain different
numbers of photon (or electron) lines, thus not satisfying a simple 
eigenvalue equation like 
(\ref{vac19}). In fact, each vacuum diagram, generated by using
the right-hand side of the recursion relation (\ref{rec07}), must be divided
by twice the number of photon lines in the diagram to obtain the correct
weight factor. This procedure is a consequence of the left-hand side of
Eq.~(\ref{rec07}), which counts the number of photon lines
in each diagram separately. By taking this into consideration, the 
second-order vacuum diagrams are given by
\begin{equation}
  \label{rec09}
W^{(2)}[J]=W^{(2)}[0]\;-
\parbox{23mm}{\begin{center}
\begin{fmfgraph}(19,6)
\setval
\fmfforce{0w,0h}{i1}
\fmfforce{0w,1h}{i2}
\fmfforce{3/19w,0.5h}{v1}
\fmfforce{6/19w,0.5h}{v2}
\fmfforce{11/19w,0.5h}{v3}
\fmfforce{14/19w,0.5h}{v4}
\fmf{double,width=0.2mm}{i1,v1,i2}
\fmf{boson}{v1,v2}
\fmf{fermion,right=1}{v2,v3,v2}
\fmf{boson}{v3,v4}
\fmfi{fermion}{fullcircle rotated 180 scaled 5/19w shifted (33/38w,0.5h)}
\fmfdot{v1,v2,v3,v4}
\end{fmfgraph}
\end{center}}-
\parbox{15mm}{\begin{center}
\begin{fmfgraph}(11,6)
\setval
\fmfforce{0w,0h}{i1}
\fmfforce{0w,1h}{i2}
\fmfforce{3/11w,0.5h}{v1}
\fmfforce{6/11w,0.5h}{v2}
\fmfforce{17/22w,5.5/6h}{v3}
\fmfforce{17/22w,0.5/6h}{v4}
\fmf{double,width=0.2mm}{i1,v1,i2}
\fmf{boson}{v1,v2}
\fmf{fermion,right=0.45}{v3,v2,v4}
\fmf{fermion,right=1}{v4,v3}
\fmf{boson}{v3,v4}
\fmfdot{v1,v2,v3,v4}
\end{fmfgraph}
\end{center}}
-\frac{1}{2}
\parbox{21mm}{\begin{center}
\begin{fmfgraph}(17,6)
\setval
\fmfforce{0w,0h}{i1}
\fmfforce{0w,1h}{i2}
\fmfforce{3/17w,0.5h}{v1}
\fmfforce{6/17w,0.5h}{v2}
\fmfforce{11/17w,0.5h}{v3}
\fmfforce{14/17w,0.5h}{v4}
\fmfforce{1w,1h}{o1}
\fmfforce{1w,0h}{o2}
\fmf{double,width=0.2mm}{i1,v1,i2}
\fmf{double,width=0.2mm}{o2,v4,o1}
\fmf{boson}{v1,v2}
\fmf{fermion,right=1}{v2,v3,v2}
\fmf{boson}{v3,v4}
\fmfdot{v1,v2,v3,v4}
\end{fmfgraph}
\end{center}}
\end{equation}
with the source-free diagrams given in (\ref{vac29}). In third order, there
are 15 diagrams which couple to the physical source:
\begin{eqnarray}
  \label{rec10}
  W^{(3)}[J]&=&W^{(3)}[0]\;-
\parbox{15mm}{\begin{center}
\begin{fmfgraph}(11,6)
\setval
\fmfforce{0w,0h}{o1}
\fmfforce{0w,1h}{o2}
\fmfforce{3/11w,0.5h}{v1}
\fmfforce{6/11w,0.5h}{i1}
\fmfforce{7.727/11w,0.897h}{i2}
\fmfforce{10.523/11w,0.745h}{i3}
\fmfforce{10.523/11w,0.255h}{i4}
\fmfforce{7.727/11w,0.103h}{i5}
\fmf{double,width=0.2mm}{o1,v1,o2}
\fmf{fermion,right=0.3}{i1,i5,i4,i3,i2,i1}
\fmf{boson}{i3,i5}
\fmf{boson}{i2,i4}
\fmf{boson}{i1,v1}
\fmfdot{i1,i2,i3,i4,i5,v1}
\end{fmfgraph}
\end{center}}-
\parbox{15mm}{\begin{center}
\begin{fmfgraph}(11,6)
\setval
\fmfforce{0w,0h}{o1}
\fmfforce{0w,1h}{o2}
\fmfforce{3/11w,0.5h}{v1}
\fmfforce{6/11w,0.5h}{i1}
\fmfforce{7.727/11w,0.897h}{i2}
\fmfforce{10.523/11w,0.745h}{i3}
\fmfforce{10.523/11w,0.255h}{i4}
\fmfforce{7.727/11w,0.103h}{i5}
\fmf{double,width=0.2mm}{o1,v1,o2}
\fmf{fermion,right=0.3}{i1,i5,i4,i3,i2,i1}
\fmf{boson}{i2,i5}
\fmf{boson,right=0.7}{i3,i4}
\fmf{boson}{i1,v1}
\fmfdot{i1,i2,i3,i4,i5,v1}
\end{fmfgraph}
\end{center}}-
\parbox{15mm}{\begin{center}
\begin{fmfgraph}(11,6)
\setval
\fmfforce{0w,0h}{o1}
\fmfforce{0w,1h}{o2}
\fmfforce{3/11w,0.5h}{v1}
\fmfforce{6/11w,0.5h}{i1}
\fmfforce{7.727/11w,0.897h}{i2}
\fmfforce{10.523/11w,0.745h}{i3}
\fmfforce{10.523/11w,0.255h}{i4}
\fmfforce{7.727/11w,0.103h}{i5}
\fmf{double,width=0.2mm}{o1,v1,o2}
\fmf{fermion,right=0.3}{i1,i5,i4,i3,i2,i1}
\fmf{boson,right=0.7}{i2,i3}
\fmf{boson,right=0.7}{i4,i5}
\fmf{boson}{i1,v1}
\fmfdot{i1,i2,i3,i4,i5,v1}
\end{fmfgraph}
\end{center}}+
\parbox{23mm}{\begin{center}
\begin{fmfgraph}(19,6)
\setval
\fmfforce{0w,0h}{s1}
\fmfforce{0w,1h}{s2}
\fmfforce{3/19w,0.5h}{s3}
\fmfforce{6/19w,0.5h}{i1}
\fmfforce{8.5/19w,0.083h}{i2}
\fmfforce{8.5/19w,0.917h}{i3}
\fmfforce{16.5/19w,0.083h}{o1}
\fmfforce{16.5/19w,0.917h}{o2}
\fmf{double,width=0.2mm}{s1,s3,s2}
\fmf{boson}{s3,i1}
\fmf{fermion,right=0.45}{i3,i1,i2}
\fmf{fermion,right=1}{i2,i3}
\fmf{boson}{i2,o1}
\fmf{boson}{i3,o2}
\fmf{fermion,right=1}{o1,o2,o1}
\fmfdot{s3,i1,i2,i3,o1,o2}
\end{fmfgraph}
\end{center}}+
\parbox{23mm}{\begin{center}
\begin{fmfgraph}(19,6)
\setval
\fmfforce{0w,0h}{s1}
\fmfforce{0w,1h}{s2}
\fmfforce{3/19w,0.5h}{s3}
\fmfforce{6/19w,0.5h}{i1}
\fmfforce{8.5/19w,0.083h}{i2}
\fmfforce{8.5/19w,0.917h}{i3}
\fmfforce{11/19w,0.5h}{i4}
\fmfforce{14/19w,0.5h}{o1}
\fmf{double,width=0.2mm}{s1,s3,s2}
\fmf{boson}{s3,i1}
\fmf{fermion,right=0.45}{i2,i4,i3,i1,i2}
\fmf{boson}{i4,o1}
\fmf{boson}{i3,i2}
\fmfi{fermion}{fullcircle rotated 180 scaled 5/19w shifted (16.5/19w,0.5h)}
\fmfdot{s3,i1,i2,i3,i4,o1}
\end{fmfgraph}
\end{center}}\nonumber\\
&&+\;
\parbox{23mm}{\begin{center}
\begin{fmfgraph}(19,6)
\setval
\fmfforce{0w,0h}{s1}
\fmfforce{0w,1h}{s2}
\fmfforce{3/19w,0.5h}{s3}
\fmfforce{6/19w,0.5h}{i1}
\fmfforce{11/19w,0.5h}{i2}
\fmfforce{14/19w,0.5h}{o1}
\fmfforce{16.5/19w,0.083h}{o2}
\fmfforce{16.5/19w,0.917h}{o3}
\fmf{double,width=0.2mm}{s1,s3,s2}
\fmf{boson}{s3,i1}
\fmf{fermion,right=1}{i1,i2,i1}
\fmf{boson}{i2,o1}
\fmf{fermion,right=0.45}{o3,o1,o2}
\fmf{fermion,right=1}{o2,o3}
\fmf{boson}{o3,o2}
\fmfdot{s3,i1,i2,o1,o2,o3}
\end{fmfgraph}
\end{center}}+
\parbox{23mm}{\begin{center}
\begin{fmfgraph}(19,6)
\setval
\fmfforce{0w,0h}{s1}
\fmfforce{0w,1h}{s2}
\fmfforce{3/19w,0.5h}{s3}
\fmfforce{6/19w,0.5h}{i1}
\fmfforce{7.25/19w,5.165/6h}{i2}
\fmfforce{9.75/19w,5.165/6h}{i3}
\fmfforce{11/19w,0.5h}{i4}
\fmfforce{14/19w,0.5h}{o1}
\fmf{double,width=0.2mm}{s1,s3,s2}
\fmf{boson}{s3,i1}
\fmf{fermion,right=0.33}{i4,i3,i2,i1}
\fmf{fermion,right=1}{i1,i4}
\fmf{boson}{i4,o1}
\fmf{boson,left=0.8}{i3,i2}
\fmfi{fermion}{fullcircle rotated 180 scaled 5/19w shifted (16.5/19w,0.5h)}
\fmfdot{s3,i1,i2,i3,i4,o1}
\end{fmfgraph}
\end{center}}+
\parbox{23mm}{\begin{center}
\begin{fmfgraph}(19,6)
\setval
\fmfforce{0w,0h}{s1}
\fmfforce{0w,1h}{s2}
\fmfforce{3/19w,0.5h}{s3}
\fmfforce{6/19w,0.5h}{i1}
\fmfforce{7.25/19w,5.165/6h}{i2}
\fmfforce{9.75/19w,5.165/6h}{i3}
\fmfforce{11/19w,0.5h}{i4}
\fmfforce{14/19w,0.5h}{o1}
\fmf{double,width=0.2mm}{s1,s3,s2}
\fmf{boson}{s3,i1}
\fmf{fermion,left=0.33}{i1,i2,i3,i4}
\fmf{fermion,left=1}{i4,i1}
\fmf{boson}{i4,o1}
\fmf{boson,left=0.8}{i3,i2}
\fmfi{fermion}{fullcircle rotated 180 scaled 5/19w shifted (16.5/19w,0.5h)}
\fmfdot{s3,i1,i2,i3,i4,o1}
\end{fmfgraph}
\end{center}}-
\parbox{31mm}{\begin{center}
\begin{fmfgraph}(27,6)
\setval
\fmfforce{0w,0h}{s1}
\fmfforce{0w,1h}{s2}
\fmfforce{3/27w,0.5h}{s3}
\fmfforce{6/27w,0.5h}{i1}
\fmfforce{11/27w,0.5h}{i2}
\fmfforce{14/27w,0.5h}{i3}
\fmfforce{19/27w,0.5h}{i4}
\fmfforce{22/27w,0.5h}{o1}
\fmf{double,width=0.2mm}{s1,s3,s2}
\fmf{boson}{s3,i1}
\fmf{fermion,right=1}{i1,i2,i1}
\fmf{boson}{i2,i3}
\fmf{fermion,right=1}{i3,i4,i3}
\fmf{boson}{i4,o1}
\fmfi{fermion}{fullcircle rotated 180 scaled 5/27w shifted (24.5/27w,0.5h)}
\fmfdot{s3,i1,i2,i3,i4,o1}
\end{fmfgraph}
\end{center}}-
\parbox{19mm}{\begin{center}
\begin{fmfgraph}(15,18.856)
\setval
\fmfforce{0w,6.428/18.856h}{s1}
\fmfforce{0w,12.428/18.856h}{s2}
\fmfforce{3/15w,0.5h}{s3}
\fmfforce{6/15w,0.5h}{i1}
\fmfforce{9.75/15w,7.263/18.856h}{i2}
\fmfforce{9.75/15w,11.593/18.856h}{i3}
\fmfforce{11.25/15w,4.665/18.856h}{o1}
\fmfforce{11.25/15w,14.191/18.856h}{o2}
\fmf{double,width=0.2mm}{s1,s3,s2}
\fmf{boson}{s3,i1}
\fmf{fermion,right=0.6}{i1,i2,i3,i1}
\fmf{boson}{i2,o1}
\fmf{boson}{i3,o2}
\fmfi{fermion}{fullcircle rotated 120 scaled 5/15w 
shifted (12.5/15w,2.5/18.856h)}
\fmfi{fermion}{fullcircle rotated -120 scaled 5/15w 
shifted (12.5/15w,16.356/18.856h)}
\fmfdot{s3,i1,i2,i3,o1,o2}
\end{fmfgraph}
\end{center}}\nonumber\\
&&-\;\frac{1}{2}
\parbox{21mm}{\begin{center}
\begin{fmfgraph}(17,6)
\setval
\fmfforce{0w,0h}{s1}
\fmfforce{0w,1h}{s2}
\fmfforce{3/17w,0.5h}{s3}
\fmfforce{6/17w,0.5h}{i1}
\fmfforce{8.5/17w,0.083h}{i2}
\fmfforce{8.5/17w,0.917h}{i3}
\fmfforce{11/17w,0.5h}{i4}
\fmfforce{14/17w,0.5h}{o1}
\fmfforce{1w,0h}{o2}
\fmfforce{1w,1h}{o3}
\fmf{double,width=0.2mm}{s1,s3,s2}
\fmf{double,width=0.2mm}{o2,o1,o3}
\fmf{boson}{s3,i1}
\fmf{fermion,right=0.45}{i2,i4,i3,i1,i2}
\fmf{boson}{i4,o1}
\fmf{boson}{i3,i2}
\fmfdot{s3,i1,i2,i3,i4,o1}
\end{fmfgraph}
\end{center}}-
\parbox{21mm}{\begin{center}
\begin{fmfgraph}(17,6)
\setval
\fmfforce{0w,0h}{s1}
\fmfforce{0w,1h}{s2}
\fmfforce{3/17w,0.5h}{s3}
\fmfforce{6/17w,0.5h}{i1}
\fmfforce{7.25/17w,5.165/6h}{i2}
\fmfforce{9.75/17w,5.165/6h}{i3}
\fmfforce{11/17w,0.5h}{i4}
\fmfforce{14/17w,0.5h}{o1}
\fmfforce{1w,0h}{o2}
\fmfforce{1w,1h}{o3}
\fmf{double,width=0.2mm}{s1,s3,s2}
\fmf{double,width=0.2mm}{o2,o1,o3}
\fmf{boson}{s3,i1}
\fmf{fermion,left=0.33}{i1,i2,i3,i4}
\fmf{fermion,left=1}{i4,i1}
\fmf{boson}{i4,o1}
\fmf{boson,left=0.8}{i3,i2}
\fmfdot{s3,i1,i2,i3,i4,o1}
\end{fmfgraph}
\end{center}}-
\parbox{19mm}{\begin{center}
\begin{fmfgraph}(15,18.856)
\setval
\fmfforce{0w,6.428/18.856h}{s1}
\fmfforce{0w,12.428/18.856h}{s2}
\fmfforce{3/15w,0.5h}{s3}
\fmfforce{6/15w,0.5h}{i1}
\fmfforce{9.75/15w,7.263/18.856h}{i2}
\fmfforce{9.75/15w,11.593/18.856h}{i3}
\fmfforce{11.25/15w,4.665/18.856h}{o1}
\fmfforce{11.25/15w,14.191/18.856h}{o2}
\fmfforce{15.35/15w,15.29/18.856h}{o3}
\fmfforce{10.15/15w,18.29/18.856h}{o4}
\fmf{double,width=0.2mm}{s1,s3,s2}
\fmf{double,width=0.2mm}{o4,o2,o3}
\fmf{boson}{s3,i1}
\fmf{fermion,right=0.6}{i1,i2,i3,i1}
\fmf{boson}{i2,o1}
\fmf{boson}{i3,o2}
\fmfi{fermion}{fullcircle rotated 120 scaled 5/15w 
shifted (12.5/15w,2.5/18.856h)}
\fmfdot{s3,i1,i2,i3,o1,o2}
\end{fmfgraph}
\end{center}}+\;\frac{1}{2}
\parbox{29mm}{\begin{center}
\begin{fmfgraph}(25,6)
\setval
\fmfforce{0w,0h}{s1}
\fmfforce{0w,1h}{s2}
\fmfforce{3/25w,0.5h}{s3}
\fmfforce{6/25w,0.5h}{i1}
\fmfforce{11/25w,0.5h}{i2}
\fmfforce{14/25w,0.5h}{i3}
\fmfforce{19/25w,0.5h}{i4}
\fmfforce{22/25w,0.5h}{o1}
\fmfforce{1w,0h}{o2}
\fmfforce{1w,1h}{o3}
\fmf{double,width=0.2mm}{s1,s3,s2}
\fmf{double,width=0.2mm}{o2,o1,o3}
\fmf{boson}{s3,i1}
\fmf{fermion,right=1}{i1,i2,i1}
\fmf{boson}{i2,i3}
\fmf{fermion,right=1}{i3,i4,i3}
\fmf{boson}{i4,o1}
\fmfdot{s3,i1,i2,i3,i4,o1}
\end{fmfgraph}
\end{center}}-\;\frac{1}{3}
\parbox{19mm}{\begin{center}
\begin{fmfgraph}(15,18.856)
\setval
\fmfforce{0w,6.428/18.856h}{s1}
\fmfforce{0w,12.428/18.856h}{s2}
\fmfforce{3/15w,0.5h}{s3}
\fmfforce{6/15w,0.5h}{i1}
\fmfforce{9.75/15w,7.263/18.856h}{i2}
\fmfforce{9.75/15w,11.593/18.856h}{i3}
\fmfforce{11.25/15w,4.665/18.856h}{o1}
\fmfforce{11.25/15w,14.191/18.856h}{o2}
\fmfforce{15.35/15w,15.29/18.856h}{o3}
\fmfforce{10.15/15w,18.29/18.856h}{o4}
\fmfforce{10.15/15w,0.565/18.856h}{o5}
\fmfforce{15.35/15w,3.565/18.856h}{o6}
\fmf{double,width=0.2mm}{s1,s3,s2}
\fmf{double,width=0.2mm}{o4,o2,o3}
\fmf{double,width=0.2mm}{o6,o1,o5}
\fmf{boson}{s3,i1}
\fmf{fermion,right=0.6}{i1,i2,i3,i1}
\fmf{boson}{i2,o1}
\fmf{boson}{i3,o2}
\fmfdot{s3,i1,i2,i3,o1,o2}
\end{fmfgraph}
\end{center}}.
\end{eqnarray}
In the following we derive a recursion relation which allows us to generate
{\it only} those vacuum diagrams which contain a coupling to the source.
\subsubsection{Recursion Relation for Vacuum Diagrams Coupled to the 
External Source}
Since we have the possibility to generate all source-free vacuum diagrams
with the help of the recursion relation (\ref{vac26}), 
we are able to set up a recursion relation
to generate only the diagrams
with source coupling. 
Inserting on the left-hand side of Eq.~(\ref{rec03}) the equation
\begin{equation}
  \label{rrj00}
  \meanJ{\hat{A}_1}=\frac{1}{e}
\frac{\delta W[J]}{\delta J_1},
\end{equation}
multiplying 
both sides with $J_1$, and performing the integral $\int_1$ yields
\begin{equation}
  \label{rrj01}
  \int_1J_1\frac{\delta W[J]}{\delta J_1}=
e^2\int_{1\cdots 6}V_{234}D_{14}J_1S_{52}S_{36}
\frac{\delta W[J]}{\delta S_{56}}+e^2\int_{12}J_1 D_{12}
J_2.
\end{equation}
On the right-hand side
we have changed the functional derivatives with respect to the kernel
$S^{-1}$ into functional derivatives with respect to the propagator $S$ using 
Eq.~(\ref{rr31}).
Inserting the decomposition (\ref{rec05b}) and utilizing the fact
that $W^{(0)}$ from 
Eq.~(\ref{gt19b}) is source-free, 
$\delta W^{(0)}/\delta J_1=0$, we find 
\begin{eqnarray}
  \label{rrj02}
&& \int_1J_1\frac{\delta W^{(1)}[J]}{\delta J_1}
+\sum\limits_{n=1}^\infty e^{2n}\int_1J_1
\frac{\delta W^{(n+1)}[J]}{\delta J_1}=
\nonumber\\
&&\hspace*{10mm}-\int_{1\cdots 4}
V_{234}D_{14}J_1S_{32}+
\int_{12}J_1D_{12}J_2+\;\sum\limits_{n=1}^\infty 
e^{2n}\int_{1\cdots 6}V_{234}D_{14}J_1
S_{52}S_{36}\frac{\delta W^{(n)}[J]}{\delta S_{56}}.
\end{eqnarray}
To lowest order, the right-hand side yields
source diagrams are
\begin{equation}
  \label{rrj03}
\tilde{W}^{(1)}[J]=\frac{1}{2}
\parbox{14mm}{\begin{center}
\begin{fmfgraph}(10,6)
\setval
\fmfforce{0w,0h}{i1}
\fmfforce{3/10w,0.5h}{i2}
\fmfforce{0w,1h}{i3}
\fmfforce{1w,0h}{o1}
\fmfforce{7/10w,0.5h}{o2}
\fmfforce{1w,1h}{o3}
\fmf{double,width=0.2mm}{i1,i2,i3}
\fmf{double,width=0.2mm}{o1,o2,o3}
\fmf{boson}{i2,o2}
\fmfdot{i2,o2}
\end{fmfgraph}
\end{center}}-
\parbox{16mm}{\begin{center}
\begin{fmfgraph}(12,6)
\setval
\fmfleft{i1,i2}
\fmfforce{1/4w,0.5h}{v1}
\fmfforce{7/12w,0.5h}{v2}
\fmf{double,width=0.2mm}{i1,v1,i2}
\fmf{boson}{v1,v2}
\fmfi{fermion}{fullcircle rotated 180 scaled 5/12w shifted (0.792w,0.5h)}
\fmfdot{v1,v2}
\end{fmfgraph}
\end{center}},  
\end{equation}
where we have used the wiggle to indicate the restriction to 
the source diagrams of $W^{(1)}[J]$ in Eq.~(\ref{rec08}).
The full functional solving Eq.~(\ref{rrj02}) consists of the terms
\begin{equation}
  \label{rrj04}
W^{(n)}[J]=W^{(n)}[0]+\tilde{W}^{(n)}[J] \,  
\end{equation}
where the source-free contributions $W^{(n)}[0]=W^{(n)}$ 
of Sect.~\ref{recrel}
represent integration constants undetermined by Eq.~(\ref{rrj02}).
Introducing a diagram for the functional derivative with
respect to the current $J$,
\begin{equation}
  \label{rrj05}
\dcurr{}{1}\quad\equiv\quad \frac{\delta}{\delta J_1}  ,
\end{equation}
the recursion relation for the vacuum diagrams with source-coupling 
Eq. (\ref{rrj02}) is 
graphically written for $n\ge 1$ as
\begin{equation}
  \label{rrj06}
  \parbox{11mm}{\centerline{  
\begin{fmfgraph*}(7,6)
\setval
\fmfleft{i1,i2}
\fmfright{o}
\fmfforce{0.5w,0.5h}{v}
\fmf{double,width=0.2mm}{i1,v,i2}
\fmf{boson}{v,o}
\fmfdot{v}
\end{fmfgraph*}
}}{\scs 1}\quad\dcurr{W^{(n+1)}[J]}{1}\quad\equiv\quad
\parbox{15mm}{\begin{center}
\begin{fmfgraph*}(10,6)
\setval
\fmfstraight
\fmfleft{i2,i1}
\fmfright{o2,o1}
\fmf{double,width=0.2mm}{i2,v1,i1}
\fmf{boson}{v1,v2}
\fmf{fermion}{v2,o1}
\fmf{fermion}{o2,v2}
\fmfdot{v1,v2}
\fmfv{decor.size=0, label=${\scs 1}$, l.dist=1mm, l.angle=0}{o1}
\fmfv{decor.size=0, label=${\scs 2}$, l.dist=1mm, l.angle=0}{o2}
\end{fmfgraph*}
\end{center}}\quad\dfermi{W^{(n)}[J]}{1}{2}.
\end{equation}
The graphical operation on the right-hand side means that an external current 
is attached through a photon line to a fermion line in all possible ways.
The iteration of this recursion relation is very simple since the right-hand
side is linear. Each diagram calculated with the right-hand side of this
equation must be divided by the number of source-coupling within the
diagram since the operation on the left-hand side counts the number of
source-couplings in the diagram. By considering Eq.~(\ref{rrj04}), one easily
reproduces the higher-order vacuum diagrams given in the Eqs. (\ref{rec09}) 
and (\ref{rec10}).
\subsection{Scattering of Electrons and Photons in the Presence of an
External Source}
Typically, an external
electromagnetic field is produced by 
a heavy particle such as a nucleus or an ion.
Quantum electrodynamical effects like pair creation, Bremsstrahlung,
and Lamb shift are caused by such electromagnetic fields. The Feynman diagrams
for the $n$-point functions associated with these processes are again 
obtained by cutting 
electron or photon lines from the just-derived vacuum diagrams. 
\subsubsection{Vacuum Polarization Induced by External Field}
The photon propagator in the presence of an external source
\begin{equation}
  \label{ph00}
  {}^\gamma G^{2}_{12}[J]=
-2\frac{\delta W[J]}{\delta D^{-1}_{12}}
\end{equation}
is found by cutting a photon line in the vacuum diagrams 
(\ref{rec08})--(\ref{rec10}):
\begin{eqnarray}
\label{ph01}
{}^\gamma G^{2,c}_{12}[J]&=&{}^\gamma G^{2,c}_{12}[0]\;+
\;e^4\;\Bigg[
\;-\parbox{17mm}{\begin{center}
\begin{fmfgraph*}(9,7)
\setval
\fmfforce{0w,0h}{s1}
\fmfforce{2.5/9w,1/2h}{s2}
\fmfforce{0w,1h}{s3}
\fmfforce{1w,0h}{i1}
\fmfforce{7.25/9w,2.3/7h}{i2}
\fmfforce{5/9w,1/2h}{i3}
\fmfforce{7.25/9w,4.7/7h}{i4}
\fmfforce{1w,1h}{i5}
\fmf{double,width=0.2mm}{s1,s2,s3}
\fmf{boson}{s2,i3}
\fmf{boson}{i4,i5}
\fmf{boson}{i1,i2}
\fmf{fermion,right=0.6}{i4,i3,i2,i4}
\fmflabel{$\scs 1$}{i5}
\fmflabel{$\scs 2$}{i1}
\fmfdot{s2,i2,i3,i4}
\end{fmfgraph*}
\end{center}}-\;({\scs 1}\leftrightarrow {\scs 2})\;
\Bigg]+{\cal O}(e^6),
\end{eqnarray}
showing polarization caused by the external field. 
\subsubsection{Lamb-Shift and Anomalous Magnetic Moment}
The important phenomena of Lamb shift and anomalous magnetic moments
are obtained from the perturbative corrections in the electron 
propagator:
\begin{equation}
  \label{ll00}
  {}^e G^{2}_{12}[J]=
\frac{\delta W[J]}{\delta S^{-1}_{21}},
\end{equation}
whose diagrams come from cutting an electron line in the vacuum diagrams
(\ref{rec08})--(\ref{rec10}). To order $e^4$, we have
\begin{eqnarray}
\label{ll01}
{}^eG^{2}_{12}[J]&=&{}^eG^{2}_{12}[0]\;+\;e^2
\parbox{17mm}{\begin{center}
\begin{fmfgraph*}(9,7)
\setval
\fmfforce{0w,0h}{s1}
\fmfforce{2/9w,1/2h}{s2}
\fmfforce{0w,1h}{s3}
\fmfforce{1w,0h}{i1}
\fmfforce{7/9w,1/2h}{i2}
\fmfforce{1w,1h}{i3}
\fmf{double,width=0.2mm}{s1,s2,s3}
\fmf{boson}{s2,i2}
\fmf{fermion}{i1,i2,i3}
\fmflabel{$\scs 1$}{i3}
\fmflabel{$\scs 2$}{i1}
\fmfdot{s2,i2}
\end{fmfgraph*}
\end{center}}+e^4\;\Bigg[
\parbox{13mm}{\begin{center}
\begin{fmfgraph}(9,9)
\setval
\fmfforce{0w,0h}{s1}
\fmfforce{2/9w,1/2h}{s2}
\fmfforce{0w,1h}{s3}
\fmfforce{1w,0h}{i1}
\fmfforce{1w,1h}{i5}
\fmf{double,width=0.2mm}{s1,s2,s3}
\fmf{fermion,tension=2}{i1,i2,i3,i4,i5}
\fmf{boson}{s2,i3}
\fmffreeze
\fmf{boson,right=0.5}{i2,i4}
\fmfdot{s2,i2,i3,i4}
\end{fmfgraph}
\end{center}}+
\parbox{13mm}{\begin{center}
\begin{fmfgraph}(9,9)
\setval
\fmfforce{0w,0h}{s1}
\fmfforce{2/9w,1/2h}{s2}
\fmfforce{0w,1h}{s3}
\fmfforce{1w,0h}{i1}
\fmfforce{1w,1h}{i5}
\fmf{double,width=0.2mm}{s1,s2,s3}
\fmf{fermion,tension=1/3}{i1,i2,i3,i4}
\fmf{fermion}{i4,i5}
\fmf{boson}{s2,i4}
\fmf{boson,left=1.3,tension=0.1}{i2,i3}
\fmfdot{s2,i2,i3,i4}
\end{fmfgraph}
\end{center}}+
\parbox{13mm}{\begin{center}
\begin{fmfgraph}(9,9)
\setval
\fmfforce{0w,0h}{s1}
\fmfforce{2/9w,1/2h}{s2}
\fmfforce{0w,1h}{s3}
\fmfforce{1w,0h}{i1}
\fmfforce{1w,1h}{i5}
\fmf{double,width=0.2mm}{s1,s2,s3}
\fmf{fermion,tension=1/3}{i2,i3,i4,i5}
\fmf{fermion}{i1,i2}
\fmf{boson}{s2,i2}
\fmf{boson,left=1.3,tension=0.1}{i3,i4}
\fmfdot{s2,i2,i3,i4}
\end{fmfgraph}
\end{center}}-
\parbox{13mm}{\begin{center}
\begin{fmfgraph}(9,9)
\setval
\fmfforce{0w,0h}{s1}
\fmfforce{2/9w,1/2h}{s2}
\fmfforce{0w,1h}{s3}
\fmfforce{1w,0h}{i1}
\fmfforce{1w,1h}{i4}
\fmfforce{0.6w,0.2h}{i5}
\fmf{double,width=0.2mm}{s1,s2,s3}
\fmf{fermion,tension=2}{i1,i2,i3}
\fmf{fermion,tension=2}{i3,i4}
\fmf{boson}{s2,i3}
\fmf{boson,tension=0}{i2,i5}
\fmfi{fermion}{fullcircle rotated 30 scaled 5/18w shifted (0.48w,0.16h)}
\fmfdot{s2,i2,i3,i5}
\end{fmfgraph}
\end{center}}-
\parbox{13mm}{\begin{center}
\begin{fmfgraph}(9,9)
\setval
\fmfforce{0w,0h}{s1}
\fmfforce{2/9w,1/2h}{s2}
\fmfforce{0w,1h}{s3}
\fmfforce{1w,0h}{i1}
\fmfforce{1w,1h}{i4}
\fmfforce{0.6w,0.8h}{i5}
\fmf{double,width=0.2mm}{s1,s2,s3}
\fmf{fermion,tension=2}{i1,i2}
\fmf{fermion,tension=2}{i2,i3,i4}
\fmf{boson}{s2,i2}
\fmf{boson,tension=0}{i3,i5}
\fmfi{fermion}{fullcircle rotated -30 scaled 5/18w shifted (0.48w,0.84h)}
\fmfdot{s2,i2,i3,i5}
\end{fmfgraph}
\end{center}}-
\parbox{14.5mm}{\begin{center}
\begin{fmfgraph}(10.5,9)
\setval
\fmfforce{0w,0h}{s1}
\fmfforce{2/10.5w,1/2h}{s2}
\fmfforce{0w,1h}{s3}
\fmfforce{1w,0h}{i1}
\fmfforce{8.5/10.5w,.5h}{i2}
\fmfforce{1w,1h}{i3}
\fmfforce{4/10.5w,0.5h}{v1}
\fmfforce{6.5/10.5w,0.5h}{v2}
\fmf{double,width=0.2mm}{s1,s2,s3}
\fmf{fermion}{i1,i2,i3}
\fmf{boson}{s2,v1}
\fmf{fermion,right=1}{v1,v2,v1}
\fmf{boson}{v2,i2}
\fmfdot{s2,i2,v1,v2}
\end{fmfgraph}
\end{center}}\nonumber\\
&&+
\parbox{13mm}{\begin{center}
\begin{fmfgraph}(9,11)
\setval
\fmfforce{0w,0h}{s1}
\fmfforce{2/9w,2.5/11h}{s2}
\fmfforce{0w,5/11h}{s2b}
\fmfforce{2/9w,8.5/11h}{s3}
\fmfforce{0w,6/11h}{s3b}
\fmfforce{0w,1h}{s4}
\fmfforce{1w,0h}{i1}
\fmfforce{7/9w,2.5/11h}{i2}
\fmfforce{7/9w,8.5/11h}{i3}
\fmfforce{1w,1h}{i4}
\fmf{double,width=0.2mm}{s1,s2,s2b}
\fmf{double,width=0.2mm}{s3b,s3,s4}
\fmf{boson}{s2,i2}
\fmf{boson}{s3,i3}
\fmf{fermion}{i1,i2,i3,i4}
\fmfdot{s2,s3,i2,i3}
\end{fmfgraph}
\end{center}}
\Bigg]+{\cal O}(e^6).
\end{eqnarray}
As already mentioned before, diagrams with corrections on external legs and tadpole graphs
do not contribute to $S$-matrix elements. In some problems, diagrams with more than one 
source-coupling are irrelevant. 
\subsubsection{Pair Creation, Pair Annihilation and Bremsstrahlung}
By differentiating the vacuum energy diagrams (\ref{rec08})--(\ref{rec10})
with respect to the interaction function $V_{123}$, we obtain the vertex 
function in the presence of an external field:
\begin{equation}
  \label{pc00}
  G^3_{123}[J]=
-\frac{1}{e}\frac{\delta W[J]}{\delta V_{213}}.
\end{equation}
The connected Feynman diagrams are to order $e^3$:
\begin{eqnarray}
\label{pc01}
G^{3,c}_{123}[J]&=&G^{3,c}_{123}[0]\;+\;e^3\;\Bigg[
\parbox{15mm}{\begin{center}
\begin{fmfgraph*}(10,6)
\setval
\fmfforce{0w,0h}{s1}
\fmfforce{2/10w,1/2h}{s2}
\fmfforce{0w,1h}{s3}
\fmfforce{3/10w,0h}{i1}
\fmfforce{1w,0h}{i2}
\fmfforce{6.5/10w,1h}{i3}
\fmfforce{5/10w,1/2h}{v1}
\fmfforce{8/10w,1/2h}{v2}
\fmf{double,width=0.2mm}{s1,s2,s3}
\fmf{boson}{s2,v1}
\fmf{boson}{i3,v2}
\fmf{fermion}{i2,v2,v1,i1}
\fmfv{decor.size=0,label={\footnotesize 1},l.dist=0.5mm}{i1}
\fmfv{decor.size=0,label={\footnotesize 2},l.dist=0.5mm}{i2}
\fmfv{decor.size=0,label={\footnotesize 3},l.dist=0.5mm}{i3}
\fmfdot{s2,v1,v2}
\end{fmfgraph*}
\end{center}}+
\parbox{15mm}{\begin{center}
\begin{fmfgraph*}(10,6)
\setval
\fmfforce{1w,0h}{s1}
\fmfforce{8/10w,1/2h}{s2}
\fmfforce{1w,1h}{s3}
\fmfforce{0w,0h}{i1}
\fmfforce{7/10w,0h}{i2}
\fmfforce{3.5/10w,1h}{i3}
\fmfforce{2/10w,1/2h}{v1}
\fmfforce{5/10w,1/2h}{v2}
\fmf{double,width=0.2mm}{s1,s2,s3}
\fmf{boson}{s2,v2}
\fmf{boson}{i3,v1}
\fmf{fermion}{i2,v2,v1,i1}
\fmfv{decor.size=0,label={\footnotesize 1},l.dist=0.5mm}{i1}
\fmfv{decor.size=0,label={\footnotesize 2},l.dist=0.5mm}{i2}
\fmfv{decor.size=0,label={\footnotesize 3},l.dist=0.5mm}{i3}
\fmfdot{s2,v1,v2}
\end{fmfgraph*}
\end{center}}
\Bigg]+{\cal O}(e^5)
\end{eqnarray}
with $G^{3,c}_{123}[0]=G^{3,c}_{123}$ of Eq.~(\ref{np18}).
These diagrams appear in
pair creation, pair annihilation, or Bremsstrahlung processes.
\section{Summary}
We have introduced a graphical recursion relation obeyed by the 
vacuum diagrams 
in quantum electrodynamics based on functional analytic methods developed in
Ref.~\cite{Kleinert1,Kleinert2}. Its iterative solution
allows us to generate all
vacuum diagrams with their correct weights order by order in perturbation
theory by removing and joining lines. 
By removing photon and electron lines as well as vertices from the vacuum graphs, 
we obtain all
diagrams of scattering processes with their 
multiplicities. The method also generates all diagrams of processes involving 
external sources. 
\section*{Acknowledgements}
We thank B.~Kastening for useful discussions, 
for a careful reading of the manuscript, and for numerous suggestions for improvements.

One of us (M.B.) is supported by the Studienstiftung des deutschen Volkes.
\end{fmffile}


\begin{thebibliography}{199}
\bibitem{comment1} For compact $A$ fields, periodic boundary conditions
would have the same effect. Twisted boundary conditions would need a
separate study.
\end{thebibliography}
\end{document}